\documentclass{osa-article}
\journal{boe}
\usepackage{xspace}
\usepackage{amsmath}
\usepackage{graphicx}
\usepackage{xcolor}
\articletype{Research Article}

\newcommand{\etal}{\textit{et al.\@}\xspace}
\newcommand{\exvivo}{\textit{ex vivo}\xspace}

\newcommand{\invivo}{\textit{in vivo}\xspace}

\newcommand{\enface}{\textit{en face}\xspace}
\newcommand{\Enface}{\textit{En face}\xspace}
\newcommand{\um}{\(\muup\)m\xspace}
\newcommand{\uM}{\(\muup\)M\xspace}

\newcommand{\sqmm}{mm$^2$\xspace}

\newcommand{\ave}[2]{\left<{#1}\right>_{#2}\xspace}
\newcommand{\cdeg}{$^\circ$C\xspace}
\newcommand{\cotwo}{CO$_2$\xspace}

\newcommand{\OCDSe}{OCDS$_e$\xspace}
\newcommand{\OCDSl}{OCDS$_l$\xspace}
\newcommand{\mpms}{$\times 10^{-4}$ ms$^{-1}$\xspace}

\graphicspath{{./}{./Figs/}}

\begin{document}
\title{Label-free drug response evaluation of human derived tumor spheroids using three-dimensional dynamic optical coherence tomography}
	
\author{%
		Ibrahim Abd El-Sadek,\authormark{1,2}
		Larina Tzu-Wei Shen,\authormark{3} 
		Tomoko Mori,\authormark{3}
		Shuichi Makita,\authormark{1}
		Pradipta Mukherjee,\authormark{1} 
		Antonia Lichtenegger,\authormark{1,4} 
		Satoshi Matsusaka,\authormark{3} 
		and Yoshiaki Yasuno,\authormark{1,*} }
	
\address{%
\authormark{1}Computational Optics Group, University of Tsukuba, Tsukuba, Ibaraki 305-8573, Japan.\\
\authormark{2}Department of Physics, Faculty of Science, Damietta University, New Damietta City 34517, Damietta, Egypt.\\
\authormark{3}Clinical Research and Regional Innovation, Faculty of Medicine, University of Tsukuba, Ibaraki 305-8575, Japan.\\
\authormark{4} Center for Medical Physics and Biomedical Engineering, Medical University of Vienna, Währinger Gürtel 18-20, 4L, 1090, Vienna, Austria.}
\email{\authormark{*}yoshiaki.yasuno@cog-labs.org}
\homepage{https://optics.bk.tsukuba.ac.jp/COG/} 
	
\begin{abstract}
We demonstrate label-free drug response evaluations of human breast (MCF-7) and colon (HT-29) cancer spheroids via dynamic optical coherence tomography (OCT). 
The MCF-7 and HT-29 spheroids were treated with paclitaxel (PTX, or Taxol) and the active metabolite of irinotecan (SN-38), respectively. 
The drugs were applied using 0 (control), 0.1, 1, and 10 \uM concentrations with treatment times of 1, 3, and 6 days. 
The samples were scanned using a repeated raster scan protocol and two dynamic OCT algorithms, logarithmic intensity variance (LIV) and late OCT correlation decay speed (\OCDSl) analyses, were applied to visualize the tissue and cellular dynamics.
Different drug response patterns of the two spheroid types were visualized clearly  and analyzed quantitatively by LIV and \OCDSl imaging. 
For both spheroid types, structural corruptions and reduction of LIV and \OCDSl were observed. These results may indicate different mechanisms of the drug action. 
The results suggest that dynamic OCT can be used to highlight drug response patterns and perform anti-cancer drug testing.
\end{abstract}     

\section{Introduction}
Cancers are ones of the deadliest human diseases \cite{Sungung2020CancerJClin,Bray2021Cancer}. 
The burden of cancer and the related mortality rate could be reduced through early detection and provision of appropriate treatment in a patient-specific manner. 
Hopefully, cancer patients may have a high chance of recovery if they are diagnosed early and treated appropriately using suitable anti-cancer drugs.
However, there are several types of cancer based on which cell type has been mutated.
The type of cancer varies from patient to patient and thus the suitable drug will also vary patient by patient.

One of the most effective approaches to understand the cancer's biology and select anti-cancer drugs appropriately is to cultivate patient-derived tumor cells as a three dimensional (3D) cluster, so-called multi-cellular tumor spheroid (MCTS).
An MCTS closely mimic the main features of \invivo solid tumors, including their structural organization and the gradients of their oxygen, pH, and nutrients \cite{Costa_Biotechnol.Advan.2016, Shahi_AssayDrugDevTechnol2019, Han_CancerCellInt2021,Lee_Biosensors_2021}.
In addition, the MCTS is similar to the \invivo solid tumor in several ways, including growth kinetics, cell interactions, metabolic rates, and its response or resistance to the applied chemotherapy and radiotherapy \cite{kobayashi_PNAS1993, Ivascu_BiomolScreen2006}.
MCTSs have therefore been used widely in the field of anti-cancer drug investigations.
An anti-cancer drug's efficacy can be evaluated based on its impact on the shape and volume alteration of the MCTS, the dissociation of the MCTS cell aggregations, and the MCTS cells viability degradation \cite{Dubois_Oncotarget2017, Thakuri_IEEE2016, Degrandis_FrontOncol2021}.

MCTS evaluations can be performed using several conventional methods, e.g., staining histology \cite{Jep2017POlSONE, plummer2019SciRep}, fluorescence microscopy \cite{pampaloni2013Cell, Mittler2017FRONTONCOL, Yang2019MaterDes}, and bright field microscopy \cite{Baek2016OncoTargetsTher, Zoetemelk_SciRep2019}.
Although these methods are the gold standards, they have several limitations. 
First, most of these methods use chemical labeling and/or tissue slicing and are thus invasive. 
Second, most of them are 2D imaging modalities.
Third, the image penetration depth of these methods is limited to a few hundreds of microns, which prevents entire-depth imaging of thick tissues such as the MCTS. 
Finally, the quantitative capabilities of these methods are low.

Some of the limitations mentioned above, e.g., invasiveness and limited penetration depth, can be overcome using optical coherence tomography (OCT) \cite{Drexler2015SprinInt} and its microscopic version so-called optical coherence microscopy (OCM) \cite{Aguirre2015OL}. 
OCT is a low-coherence interferometry-based biomedical imaging modality that provides high-resolution, label-free, non-invasive, 3D imaging over a few millimeters penetration depth.

Several hardware and software extensions have been integrated with conventional OCT to provide additional contrasts to the OCT.
For example, attenuation coefficient (AC) imaging\cite{Vermeer2014BOE, Gong2020JBO} is sensitive to tissue density, and was found to be sensitive to both the apoptotic and necrotic  processes of human fibroblasts \cite{VanDerMeer_2010}.
The tissue polarization properties can be measured using polarization-sensitive OCT \cite{DeBoer2017BOE}, which provides collagen-sensitive \cite{Sugiyama2015BOE, Li2017BOE} and melanin-sensitive \cite{Gotzinger2008OE, Makita2014OL, Miura_SciRep2021} contrasts. 
Tissue stiffness, which is an important parameter in tumor investigations, can also be evaluated via optical coherence elastography \cite{Kennedy2014BOE,kennedy_SciRep2015,Gubarkova_BOE2019, Miyazawa2019BOE}.
 
One interesting OCT extension is dynamic OCT. 
As an emerging technique, dynamic OCT provides tissue activity/sub-cellular motion contrasts.
Dynamic OCT represents a combination of high-speed OCT systems with the signal processing of sequentially acquired OCT frames. 
Time-frequency analysis-based \enface dynamic full field OCT (DFF-OCT) has been demonstrated previously and was found to be sensitive to the adenosine triphosphate (ATP) consumption-related tissue metabolism \cite{Apelian2016BOE, Thouvenin2017ApplSci, Scholler2020LSA}. 
High-speed scanning OCT and time spectroscopic analysis have been used to perform for cross-sectional dynamic imaging\cite{leung2020BOE, El-Sadek_BOE2020}. 
Both of these methods are able to provide cross-sectional tissue activity/sub-cellular motion contrast.
However, the methods may not be suitable for volumetric imaging applications.

Recently, 3D dynamic OCT methods have been demonstrated \cite{Scholler_LightSciAppl_2020, Munter2020OL, Kurokawa2020Neuro,Munter2021BOE}.
However, acquisition of the volumetric dynamics requires long measurement times \cite{Scholler_LightSciAppl_2020, Munter2020OL} or use of ultra-fast OCT systems \cite{Kurokawa2020Neuro, Munter2020OL, Munter2021BOE}. 

Recently, we demonstrated 3D dynamic OCT, which combines a standard-speed swept-source OCT device with a custom-designed 3D scanning protocol and was shown to acquire volumetric tissue dynamics in 52.4 s \cite{ElSadek_SPIE2021, El-Sadek_BOE_2021}.
The 3D tissue dynamics were visualized and quantified using our previously proposed dynamic OCT signal analysis algorithms\cite{El-Sadek_BOE2020}. 

This 3D dynamic OCT technique visualized the viable and necrotic cell layers of MCTSs successfully. 
In addition, similarity between the fluorescence microscopy and \enface dynamic OCT contrasts was demonstrated \cite{El-Sadek_BOE_2021}. 
In the same work, a preliminary study of the tumor spheroid's response to anti-cancer drug was presented. 
However, that study involved only one type of tumor (MCF-7 cell line, human breast cancer), and only one anti-cancer drug, paclitaxel (PTX, or Taxol), was used.

In general, treatment of different tumor cell types with different anti-cancer drugs leads to different spatial patterns of the anti-cancer drug responses. 
The spatial pattern differences among the different tumor spheroids and various drug types are worth to be investigated to aid in the selection of appropriate anti-cancer drugs, and these differences can be highlighted using our proposed dynamic OCT method.

In this paper, more generalized drug response studies based on the 3D dynamic OCT are presented.
Breast and colon cancers and the combination of these cancers are among the most prevalent causes of cancer-related death worldwide\cite{Newschaffer_Lancet2001, Rodriguez_BiomedCompMeth2013, Vakili_HematolOncolStemCellRes2014, Majid_BiomedCOMPUTMETH2014}.
Therefore, we selected two tumor cell types for this study, including the human breast cancer (MCF-7) and human colon cancer (HT-29) cell-lines.
The breast and colon cancer spheroids are treated using an anti-cancer drug (PTX) and an active metabolite of Irinotecan (SN-38), respectively.
Image-based 3D observation of the time-course drug response shows different spatial patterns of the drug effects among the different tumor and drug types.
In addition, quantification of the dynamic OCT and the spheroid morphology also reveals different time courses among the cell and drug types.

\section{Methods}
\label{sec:Methods}
\subsection{Protocols and samples of drug response studies}
\label{sec:DrugResponseStudyDesign}
\begin{figure}
	\centering
	\includegraphics{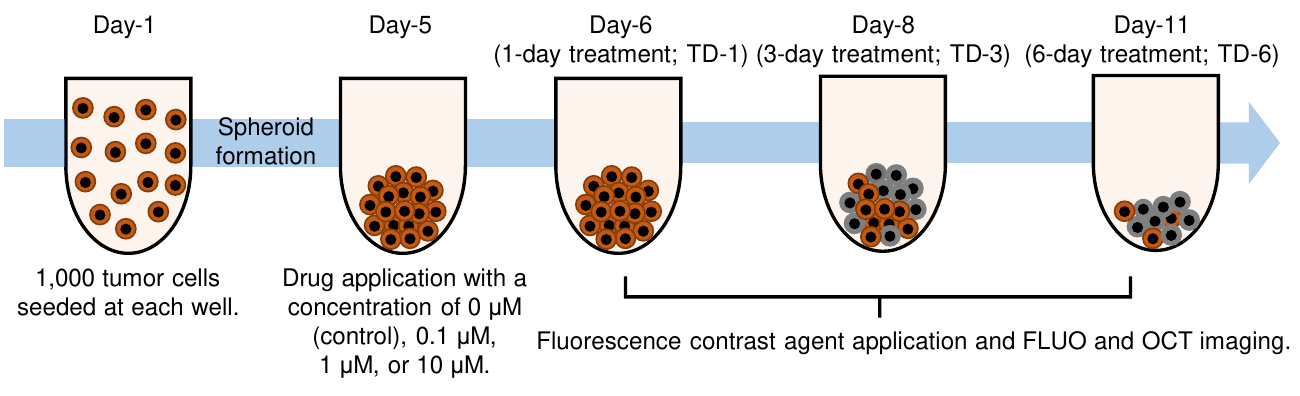} 
	\caption{Time protocol for spheroid cultivation and measurement.
		The same time protocols were used for both MCF-7 and HT-29. 
		60 spheroids were cultured for both cell types.
		Each spheroid was seeded with 1,000 cells on day-1.
		The drugs, i.e., PTX for MCF-7 and SN-38 for HT-29, were applied on day-5.
		On days-6, 8, and 11, fluorescence (FLUO) and then OCT imaging were performed.
		The fluorescence contrast agents were applied 3 hours before FLUO imaging.
	}
	\label{fig:Study_TimeCahrt}	
\end{figure}
The tumor spheroid drug response evaluation study involves use of two different human-derived tumor cell-lines, including human breast adenocarcinoma (MCF-7) and human colon cancer (HT-29) cell-lines.  
Both cell-lines were cultured using the same protocol in line with the study protocol depicted in Fig.\@ \ref{fig:Study_TimeCahrt}.
On day-1, 1,000 tumor cells were seeded in each well of a ultra-low attachment 96-well-plate and 60 wells were prepared. 
The cells were cultured at a temperature of 37 \cdeg and the cell-culture chamber was supported with 5 \% \cotwo.
The culture medium contained a 1:1 mixture of Eagle’s minimal essential medium (EMEM; for MCF-7) or Dulbecco's modified eagle medium (DMEM; for HT-29) and F12 (Invitrogen, Waltham, MA) with a 2 \% B-27 supplement (Invitrogen), 2 ng/mL of basic fibroblast growth factor (bFGF; Wako, Osaka, Japan), 2 ng/mL of epidermal growth factor (EGF; Sigma-Aldrich, St. Louis, MO), 100 U/mL of penicillin G, and 0.1 mg/mL of streptomycin sulfate (Wako, Osaka, Japan). 
The cells were aggregated with each other and formed one spheroid in each well on day-5. 
On the same day, each spheroid was then treated using 0.1 \uM, 1 \uM, or 10 \uM of the relevant anti-cancer drugs.
The MCF-7 spheroids were treated using paclitaxel (PTX, Sigma-Aldrich; also known as Taxol), while the HT-29 spheroids were treated using an active metabolite of Irinotecan (SN-38, Chemscene LLC, NJ). 
In addition, untreated (0 \uM) spheroids were retained as control samples.
After drug application the spheroids were kept within the same culture environment.

The spheroids were then extracted from the cultivation and measured using the fluorescence (FLUO) and dynamic OCT microscopes at three time-points of days-6, 8, and 11, i.e., 1-, 3-, and 6-days after drug application, respectively.
Here, we also denote the days 6, 8, and 11 as the 1st, 3rd, and 6th days of treatment (TD-1, -3, and -6). 
At each time-point, five spheroids were measured for each of the four drug concentrations, including control (0 \uM), meaning that 20 spheroids were measured in total.
The measurements were performed in a room temperature of 25 \cdeg. 

\subsection{Fluorescence microscopic imaging}
\label{sec: flouresence}
Live/dead fluorescence assays were performed using the THUNDER imaging system (Leica Microsystems, Wetzlar, Germany) with a microscopic objective with a numerical aperture (NA) of 0.12.
Two contrast agents were applied for three hours before the FLUO imaging.
The first agent is calcein acetoxymethyl (Calcein-AM; Dojindo, Kumamoto, Japan), which stains the living (viable) cells and emits a green fluorescence signal.
The second agent is propidium iodide (PI; Dojindo), which contrasts the dead cells by emitting red fluorescence signal.

\subsection{3D dynamic OCT imaging}
\label{sec:device and protocol}
\subsubsection{OCT device}
A polarization-sensitive Jones matrix swept-source OCT (JM-OCT) was used to perform the three-dimensional (3D) tissue dynamics imaging.
A light source with a central wavelength of 1.3 \um and a scanning speed of 50,000 A-lines/s was used.
The axial (in tissue) and lateral resolutions were 14 \um and 18.1 \um, respectively.
The complete specification for the JM-OCT has been published elsewhere \cite{Li2017BOE}.
Although this device is polarization-sensitive, only polarization-insensitive OCT images, which are the average of four OCT intensity images acquired through four polarization channels, were used for the tissue dynamics imaging.

\subsubsection{OCT data acquisition protocol}
\label{sec:scanProtocol}
The spheroid samples were transferred to our 3D JM-OCT system after FLUO imaging. 
For the 3D dynamics imaging, the lateral imaging field was divided into eight regions. 
Each region consists of 16 B-scan locations and was scanned repeatedly using a quick raster scanning protocol for 32 times in 6.55 s. 
Therefore, 32 sequential frames were obtained at each location in the sample with a frame repetition time of 204.8 ms.
The OCT volume consists of a total of 128 locations and was captured in 52.4 s.
The scanning area was 1 $\times$ 1 \sqmm. 
Further details about the 3D dynamics scanning protocol are available elsewhere in the literature \cite{El-Sadek_BOE_2021}.

\subsubsection{Dynamic OCT algorithm: Logarithmic intensity variance (LIV)}
\begin{figure}
	\centering
	\includegraphics{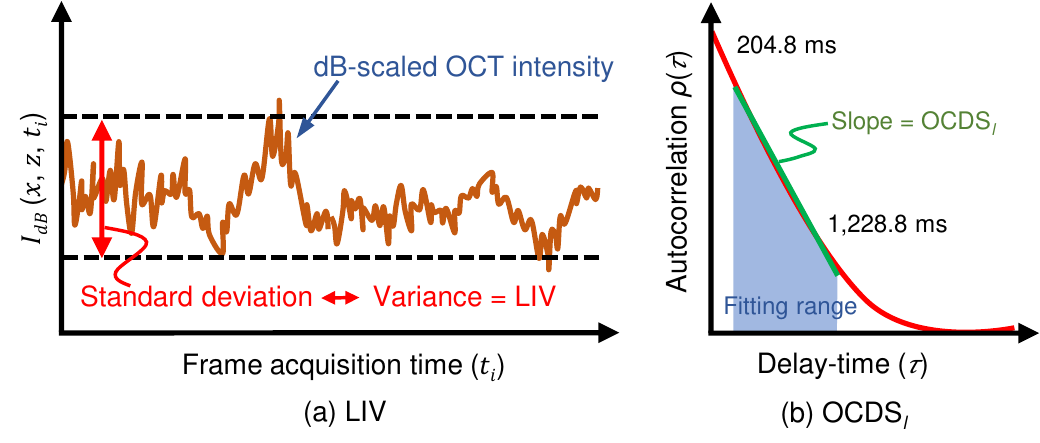} 
	\caption{Schematic depictions of two dynamic OCT algorithms:(a) LIV and (b) \OCDSl.
		LIV is defined as the temporal fluctuation (variance) of the dB-scaled OCT intensity.
		As a result, LIV is expected to be sensitive to the magnitude of the tissue dynamics.
		On the other hand, \OCDSl represents the speed of the dynamics and is defined as the decorrelation rate, i.e., as the slope of the auto-correlation curve [red curve in (b)] of the dB-scaled OCT intensity.
		The slope is computed here by performing a linear regression of the auto-correlation curve over a specified delay-time range [204.8, 1,228.8 ms]. 
	}
	\label{fig:D-OCT_Schematic}	
\end{figure}

The time-sequential OCT signals were processed using logarithmic-intensity-variance (LIV) algorithm\cite{El-Sadek_BOE2020}.
The LIV is a measure of the magnitude of the temporal fluctuations of the dB-scaled OCT signals at each location in the tissue as illustrated in Fig.\@ \ref{fig:D-OCT_Schematic}(a), and it is defined as:  
\begin{equation}
\mathrm{LIV}(x,z) = 
\frac{1}{N} \sum_{i = 0}^{N-1}
\left[I_{dB}(x,z,t_i) - \ave{I_{dB}(x,z)}{t} \right]^2\\
\label{eq:LIV}
\end{equation}
where $I_{dB}(x,z,t_i)$ is the dB-scaled (logarithmic) OCT intensity,  $t_i$ is the sampling time for the $i$-th frame, where $i$ = 0, 1, 2,...., $N$-1, and $N$ is the total number of frames, which is 32 in our case, and $\ave{}{t}$ represents averaging over time.

Because the LIV is a measure of the magnitude of the signal fluctuations, it is believed to be sensitive to the magnitude of the intracellular motility.

An LIV image is presented as a pseudo-color image in which the hue and the value of the HSV (hue, saturation, value) color representation are set as the LIV and the averaged OCT intensity image, respectively.

\subsubsection{Dynamic OCT algorithm: Late OCT correlation decay speed (\OCDSl)}
The time-sequential OCT signals are also processed using yet another dynamic OCT algorithm: ``late OCT correlation decay seed (\OCDSl).'' 
\OCDSl is defined as the speed (rate) of auto-correlation decay of the sequentially captured dB-scaled OCT intensity and it is expected to be sensitive to the slow tissue dynamics \cite{El-Sadek_BOE2020}. 

To obtain the \OCDSl, the temporal auto-correlation $\rho_A(\tau_i; x, z)$ was first computed as:
\begin{equation}
\rho_A(\tau_i; x, z) =
\frac{\mathrm{Cov}\left[I_{dB}(x,z,t_i),\, I_{dB}(x,z,t_i+\tau_i)\right]}
{\mathrm{Var}\left[I_{dB}(x,z,t_i)\right] \mathrm{Var}\left[I_{dB}(x,z,t_i+\tau_i)\right]},
\label{eq:corr_coeff}
\end{equation}
where the numerator represents the covariance between $I_{dB}(x,z,t_i)$ and $I_{dB}(x,z,t_i+\tau_i)$, and $\mathrm{Var}[\,\,]$ represents the variance. 
$\tau_i$ represents the delay time, which is defined as $i \Delta t$ where $i$ is an integer variable.
$\Delta t$ is the B-scan repetition time, which is 204.8 ms in this case. 

The \OCDSl is then defined as the correlation decay rate, i.e., as the slope of $\rho_A(\tau_i; x, z)$ over a specific range of $\tau$ = [204.8, 1228.8 ms] as illustrated in Fig.\@ \ref{fig:D-OCT_Schematic}(b).

An \OCDSl image is presented as a pseudo-color image, which is generated in the same manner as the LIV image, but in this case the hue is the \OCDSl signal.
Further details about the LIV and \OCDSl algorithms are published elsewhere \cite{El-Sadek_BOE_2021}.

\subsubsection{OCT-based volumetric quantification}
\label{sec:quantification}
To quantify the volumetric alterations of the morphology and viability of the spheroid cells, the spheroid volume was computed via B-scan by B-scan segmentation.
The segmentation was performed using the find-connected-region plugin of the Fiji ImageJ software with a manually defined OCT intensity threshold. 
The initial segmentation contained the bottom surface of the well-plate because it shows high OCT intensity.
This well-plate surface was removed manually.

In addition to the spheroid volume, the mean values of the LIV and \OCDSl over the entire spheroid volume were computed using the segmentation results.
Furthermore, by applying empirically defined cut-offs of 3 dB$^2$ and 2 \mpms for the LIV and \OCDSl, respectively, the necrotic cells volume (i.e., the volume of the LIV and \OCDSl signals that are lower than the cut-offs in each case) is computed \cite{El-Sadek_BOE_2021}.
The necrotic cell ratio (= necrotic volume / entire spheroid volume) is then computed.
These computed quantities are plotted as a function of the treatment time for each drug concentration, as shown in Section \ref{sec:res}.

\section{Results}
\label{sec:res}
\subsection{MCF-7 spheroid response to PTX}
\label{sec:MCF-7DrugResponse}
\begin{figure}
	\centering
	\includegraphics{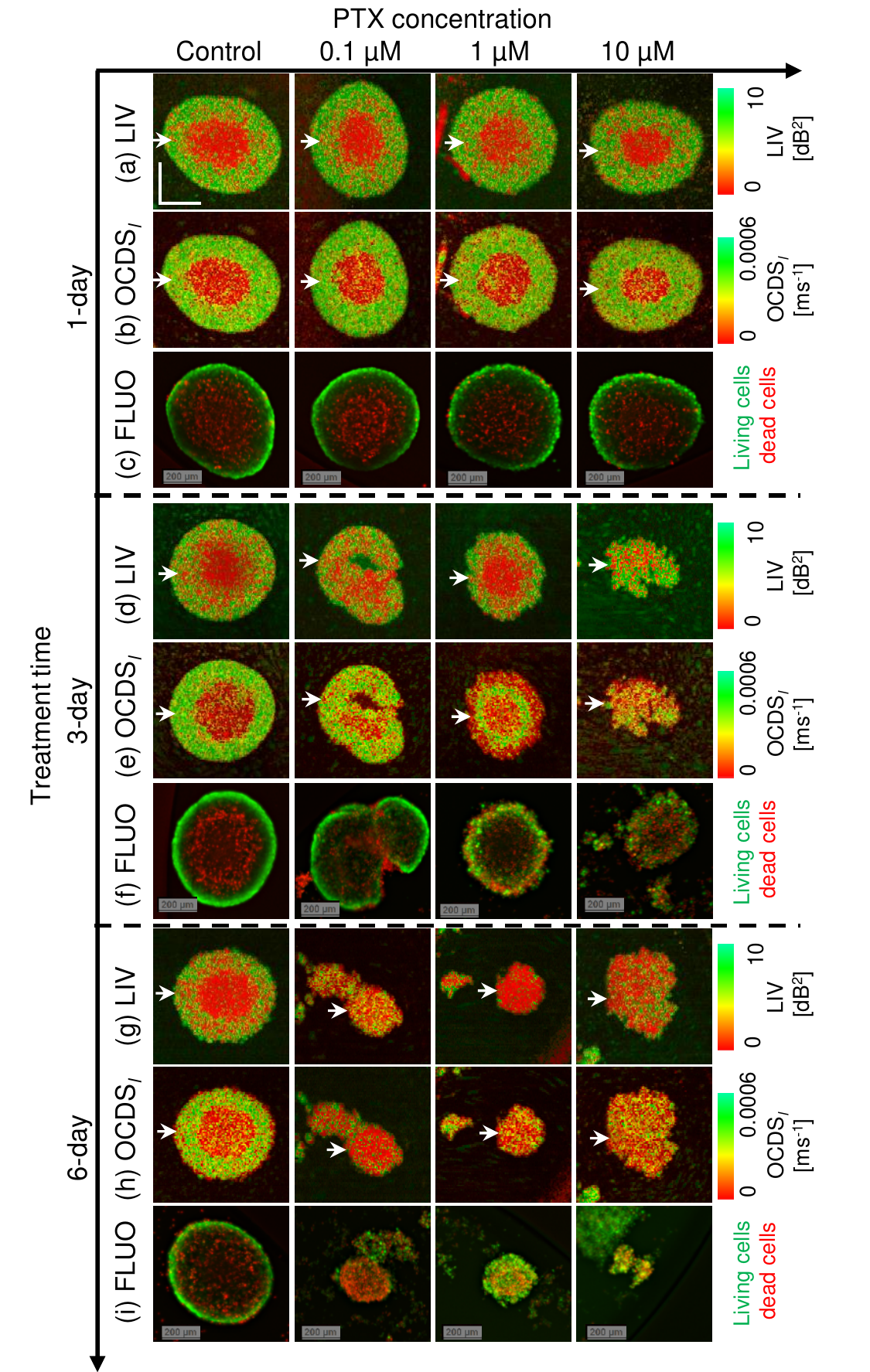} 
	\caption{LIV, \OCDSl, and FLUO images of MCF-7 spheroid treated with PTX.
		The images are \enface slices at around the center of the spheroids.
		The B-scan cross-sections and the intensity OCT images are available in the Supplementary materials (as Fig.\@ S2 and Fig.\@ S1, respectively.)
		Although the difference between the control sample and the PTX-treated samples is not obvious with the 1-day treatment, the 3- and 6-day-treated spheroids showed significantly different appearances to the control sample.
		The appearances of the FLUO images are almost consistent with those of the LIV and \OCDSl images.
		The scale bars represent 200 \um in each case.
	}

	\label{fig:MCF-7_Taxol_Image}	
\end{figure}

Figure \ref{fig:MCF-7_Taxol_Image} summarizes the \enface LIV, \OCDSl, and FLUO images of the MCF-7 spheroids treated with PTX. 
In the control case (the first column), the low LIV (red) and low \OCDSl (red) regions increase over time.
These low-dynamics regions at the center of the spheroid may correspond to the dead cells (red) shown in the FLUO images. 
It is also found that the boundaries between the high and low signal regions are clearer in the \OCDSl images than the LIV images.

On the first day of treatment (TD-1), the PTX-treated spheroids show similar appearances to the control sample.
However, as the treatment time increases, the spheroids show shape corruption, volume reduction, and degradation of the LIV and \OCDSl signals.
The FLUO images show increase of dead cells over time in addition to the shape corruption.

\begin{figure}
	\centering
	\includegraphics{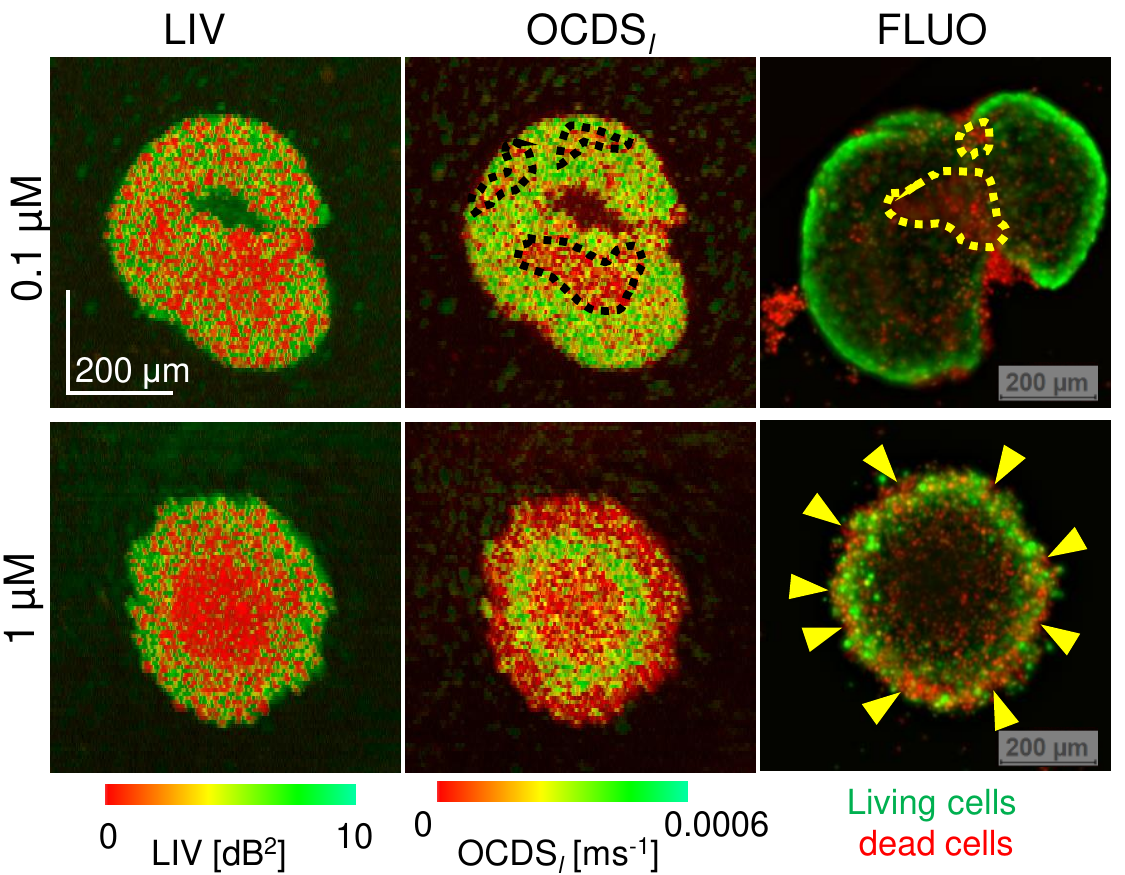} 
	\caption{Magnified LIV, \OCDSl, and FLUO images of representative spheroids from Fig.\@ \ref{fig:MCF-7_Taxol_Image}.
		The upper row shows one treated with 0.1-\uM PTX, while the lower row shows one treated with 1-\uM PTX.
		The treatment time in both cases is 3 days.
		The appearance of the FLUO images is more consistent with the \OCDSl images than LIV images.
	}
	\label{fig:MCF-7_Taxol_MagnifiedImage}	
\end{figure}
Notably, the LIV and \OCDSl images of the 0.1-\uM and 1-\uM cases on day-8 (TD-3) show completely different patterns (as magnified in Fig.\@ \ref{fig:MCF-7_Taxol_MagnifiedImage}). 
In the 0.1-\uM case, tessellated low dynamics (red) patterns appear in the \OCDSl image (black dotted shapes), while the LIV images shows  low dynamics (red) overall.
The FLUO image also shows some domains composed of dead cells (red, yellow dashed lines).
In the 1-\uM case, the LIV image shows two domains including a central low-dynamics (red) region and a peripheral region that is a mixture of low (red) and high (green) dynamics. 
In contrast, the \OCDSl image shows three concentric domains of low, high, and low dynamics.
The FLUO image shows dead cells (red) at the periphery (indicated by yellow arrowheads) and the center regions,with live cells (green) located in between.
In both cases, the appearance of the FLUO images are more consistent with \OCDSl than LIV images.

For the reference of the readers, OCT intensity images corresponding to the images shown in Fig.\@ \ref{fig:MCF-7_Taxol_Image} are presented in the supplementary material (Fig.\@ S1).
The fast-scan cross-sections of LIV and \OCDSl extracted at the locations indicated by the white arrow heads in Fig.\@ \ref{fig:MCF-7_Taxol_Image} are also presented in a supplementary figure (Fig.\@ S2). 

Images of additional spheroids measured at under each treatment condition and at each time point are presented in a supplementary figure (Fig.\@ S3) and are consistent with the images in Fig.\@ \ref{fig:MCF-7_Taxol_Image}. 

\begin{figure}
	\centering
	\includegraphics{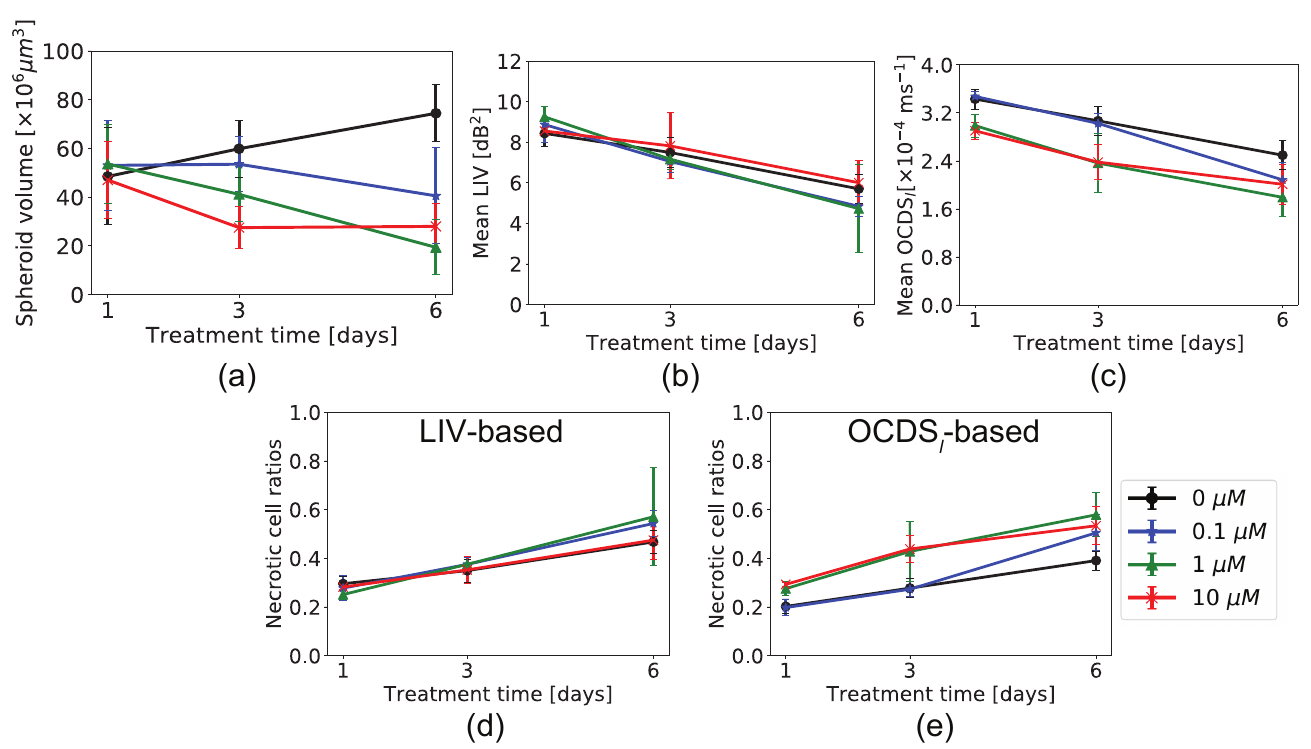}
	\caption{Quantitative analysis of the MCF-7 spheroids treated with several concentrations of Taxol. 
		(a) Spheroid volume, (b) mean LIV, (c) mean \OCDSl, and (d) and (e) necrotic cell ratios computed based on LIV and \OCDSl, respectively.
		The plotted values represent the averages of five samples in each case, while the error bars represent the $\pm$ standard-deviation range.
	}
	\label{fig:MCF7_Quant}
\end{figure} 

In Fig.\@ \ref{fig:MCF7_Quant}, the spheroid volume (a), the mean LIV (b), the mean \OCDSl (c), the LIV-based necrotic cell ratio (d), and the \OCDSl-based necrotic cell ratio (e) are plotted as functions of the treatment time.
The spheroid volumes of the control case increases over time may be due to the growth of the spheroids.
In contrast, the drug-treated spheroids show decreasing volume trends.

The mean LIV and mean \OCDSl over the entire spheroid [Fig.\@ \ref{fig:MCF7_Quant}(b) and (c)] show decreasing trends over time for all PTX concentrations, including the control case.
A significant reduction in the mean LIV was found for all PTX concentrations, except for 10\uM from TD-1 to TD-3 (P = 0.047, 0.018, and 0.006 for the control, 0.1, and 1.0 \uM) and from TD-3 to TD-6 (P = 0.006, 0.006, and 0.018 for the control, 0.1, and 1.0 \uM).
For the 10-\uM samples, the TD-1-to-TD-3 and TD-3-to-TD-6 reductions were insignificant, but TD-1-to-TD-6 reduction was significant (P = 0.006).
Significant reductions in the mean \OCDSl were also found for the three PTX concentrations from TD-1 to TD-3 (P = 0.030, 0.006, 0.018 for the control, 0.1, and 1 \uM) and from TD-3 to TD-6 (P = 0.006, 0.006, 0.030 for control, 0.1, and 1 \uM).
For the 10-\uM case, the reduction from TD-1 to TD-3 (p = 0.006) was significant, although it was not significant from TD-3 to TD-6 (p = 0.148). 
This may occur because the mean \OCDSl has already become very low on TD-3.
All the p-values of these tests are summarized in Table S1 (supplementary).

The necrotic cell ratios obtained with both LIV and \OCDSl [Fig.\@ \ref{fig:MCF7_Quant}(d) and (e), respectively] clearly increase over the treatment times for all PTX concentrations.
Significant increases in the LIV-based necrotic cell ratios were found from TD-1 to TD-3 (P = 0.030, 0.006, and 0.047 for 0.1 \uM, 1 \uM and 10 \uM, respectively), and from TD-3 to TD-6 (P = 0.006, 0.047, and 0.030 for 0.1 \uM, 1 \uM, and 10 \uM, respectively) for all PTX concentrations except for the control sample.
For the control sample, the observed increase was significant from TD-3 to TD-6 (P = 0.006).
A significant increase in the necrotic cell ratio was also found in the \OCDSl-based necrotic cell ratio from TD-1 to TD-3 for all PTX concentrations, including the control (P = 0.030, 0.018, 0.010, and 0.006 for the control, 0.1 \uM, 1 \uM and 10 \uM).
Similar increases were also found from TD-3 to TD-6 for all concentrations, except for the 10-\uM case (P = 0.006, 0.006, and 0.030 for the control,  0.1 \uM, and 1 \uM).
All statistics are summarized in Table S1 (supplementary).

The degradation in the mean LIV and mean \OCDSl and the increasing necrotic cell ratios for the PTX-treated spheroids may be related to inhibition of the intracellular transport through the microtubules.
This point will be discussed in detail later in Section \ref{sec:Diss_MCF7_Taxol}.

\subsection{HT-29 spheroid response to SN-38}
\label{sec:HT-29DrugResponse}

\begin{figure}
	\centering
	\includegraphics{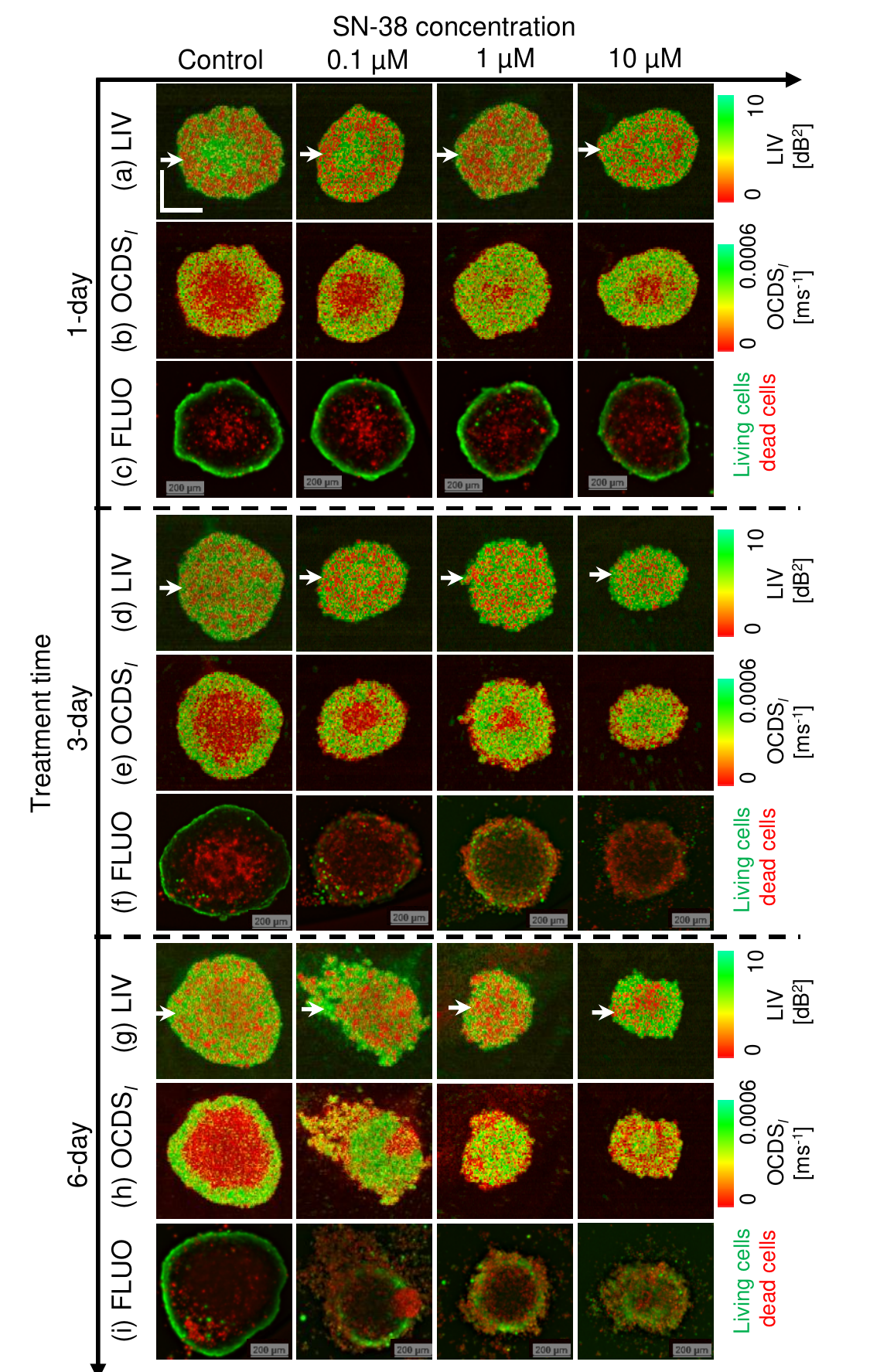} 
	\caption{\Enface LIV, \OCDSl, and FLUO images of an HT-29 spheroid treated with SN-38.
		The LIV and \OCDSl patterns are very different to those for MCF-7 when treated with PTX (Fig.\@ \ref{fig:MCF-7_Taxol_Image}).
		Specifically, the LIV images of the TD-1 spheroids show high LIV (green) at the center surrounded by low LIV (red).
		After longer treatment times, the LIV images show a granular appearance.
		The \OCDSl images show concentric domain structures.
		In addition, the \OCDSl and FLUO images are found to be highly correlated.
		The scale bars represent 200 \um in each case.}
	\label{fig:HT-29_SN38_Image}	
\end{figure} 

Figure \ref{fig:HT-29_SN38_Image} summarizes the \enface LIV, \OCDSl, and FLUO images of the HT-29 spheroids after treatment with SN-38. 
For the TD-1 spheroids at all SN-39 concentrations, including the 0 \uM (control) case, the LIV appearance is contrary to that of the MCF-7 spheroids with PTX.
Specifically, the center region shows high LIV (green), and this area is surrounded by low LIV (red) regions. 
For the 10-\uM case in particular, the low LIV region is further surrounded by a moderately high LIV periphery.
In contrast, the appearance of the \OCDSl images is similar to that in the MCF-7 case.
Namely, the center shows low \OCDSl (red), and this area is surrounded by a high \OCDSl periphery (green).

For the longer treatment times (TD-3 and TD-6), the LIV images show granular patterns of high and low LIV, and clear domain structures are not observed.
This appearance is contrary to that of the MCF-7, and is cell-type dependent.
In contrast, most of the \OCDSl images show double or triple concentric domain structures.
These structures may be the result of two factors, which include drug effect from the peripheral side and hypoxia or lack of nutrients starting from the center region.
The appearances of the \OCDSl images are quite consistent with the FLUO images.
The dynamic OCT and FLUO images show clear volume reductions and/or structural corruption under the SN-38 treatment.

\begin{figure}
	\centering
	\includegraphics{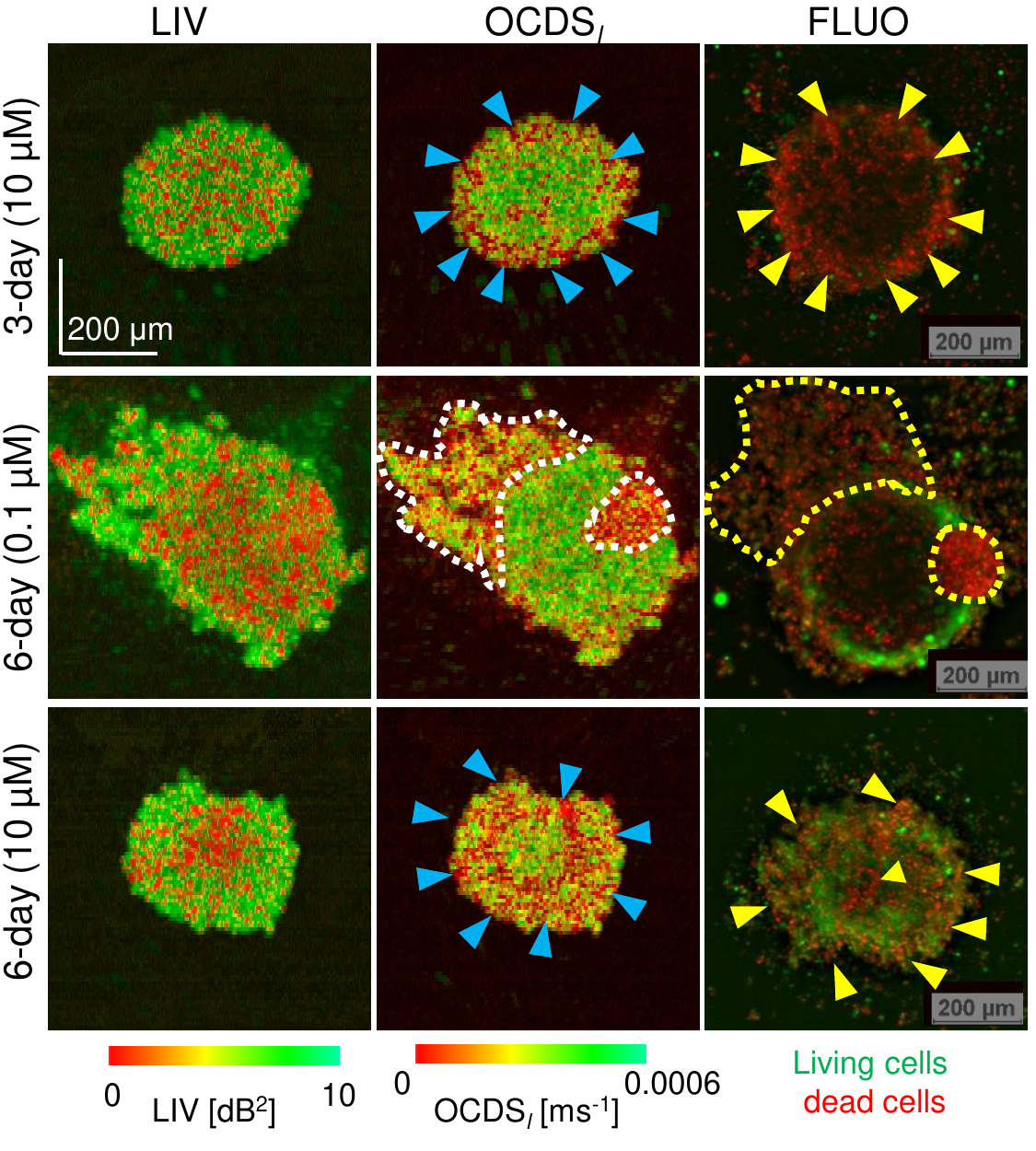} 
	\caption{Magnified images of HT-29 spheroids with sever volume reduction and shape corruption. 
		These images were extracted from Fig.\@ \ref{fig:HT-29_SN38_Image}. 
		High levels of pattern correlation between the \OCDSl and FLUO patterns are observed, while the LIV patterns are not correlated well with the other images.}
	\label{fig:HT29_SN38_MagnifiedImage}	
\end{figure}
Figure \ref{fig:HT29_SN38_MagnifiedImage} shows magnified images of representative spheroids with both volume reduction and structural corruption.
These magnified images demonstrate the high pattern correlation between the \OCDSl and FLUO images, as indicated by the arrowheads and dashed lines.
In contrast, the LIV patterns are not similar to the \OCDSl and FLUO images. 

\begin{figure}
	\centering
	\includegraphics{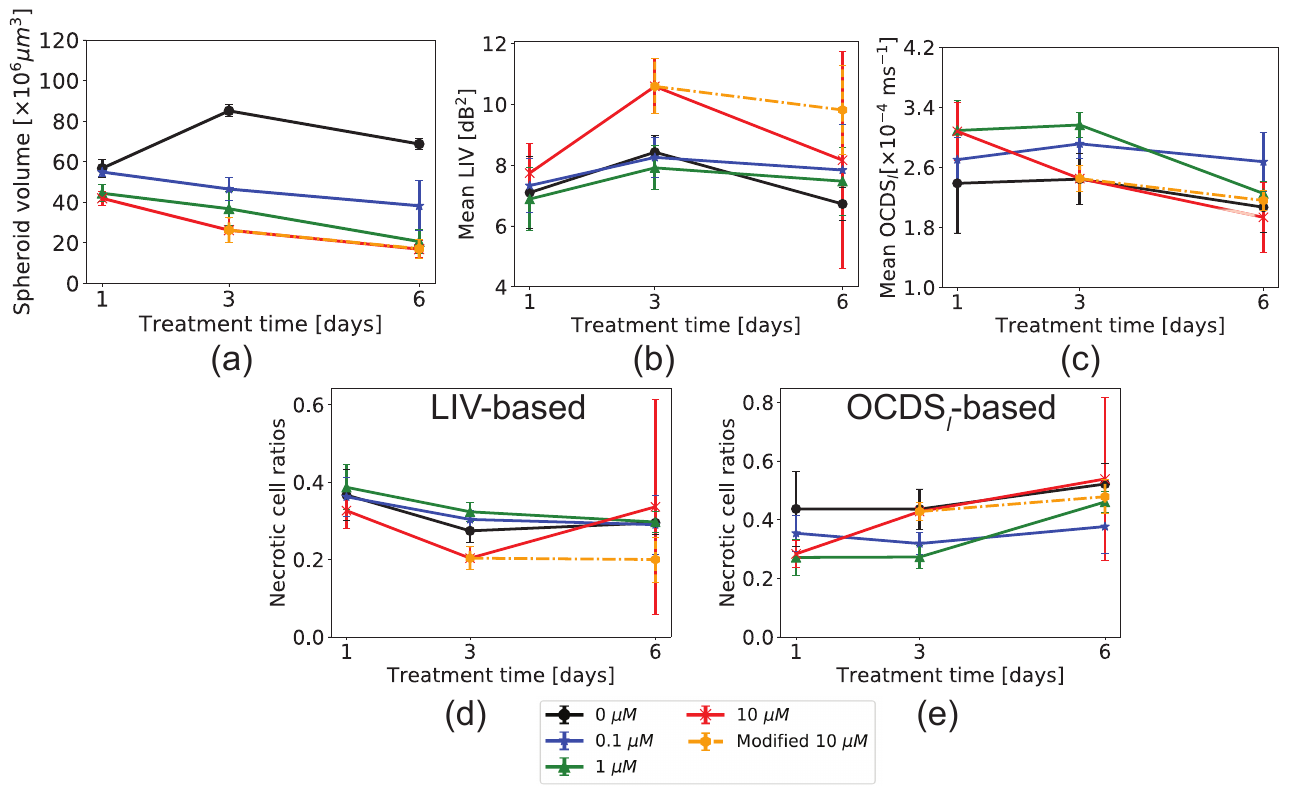}
	\caption{Morphological and dynamics analysis of the HT-29 spheroid when treated with SN-38.
		The plots are presented in the same order as those in Fig.\@ \ref{fig:MCF7_Quant}.
		The plotted values represent the average values of five samples (N=5) under each treatment condition and the error bars represent the $\pm$ standard-deviation range.
	}
	\label{fig:HT29_Quant}
\end{figure} 

\begin{figure}
	\centering
	\includegraphics{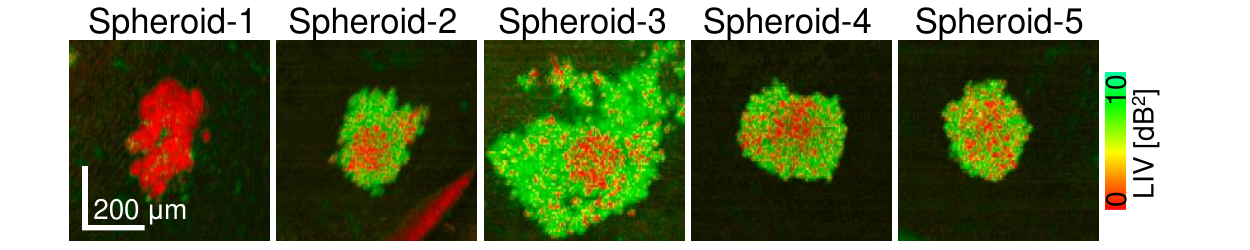}
	\caption{LIV images of the five individual spheroids when treated with 10 \uM of SN-38 for 6 days.
		The first spheroid showed exceptionally low LIV for all spheroid regions and was thus excluded from the statistical analysis as an outlier.
	}
	\label{fig:HT29_individuals}
\end{figure} 
Figure \ref{fig:HT29_Quant} shows the morphological and mean dynamic OCT signal alterations for the HT-29 spheroid treated with SN-38. 
It was observed that the data in the TD-6 10-\uM case had much standard deviations (i.e., error bars in the figure) than the other cases for the mean LIV [Fig.\@ \ref{fig:HT29_Quant}(b)], LIV-based, and \OCDSl-based necrotic cell ratios [Fig.\@ \ref{fig:HT29_Quant} (d) and (e), respectively].
By observing the LIV images for all five spheroids in this case [Fig.\@ \ref{fig:HT29_individuals}], it was found that one spheroid case (spheroid 1) exhibited very low LIV values over the entire spheroid region. 
We suspect that this spheroid may be fragmented because of some cultivation issues or because of spheroid handling during the measurement. 
We therefore regard this case as an outlier and excluded it from the subsequent statistical analysis.
The plots that include this outlier are presented with red spots and red lines, while those that exclude it are shown with orange color in Fig.\@ \ref{fig:HT29_Quant}. 

The control spheroid volume [Fig.\@ \ref{fig:HT29_Quant}(a)] increased from TD-1 to TD-3, and then decreased slightly from TD-3 to TD-6.
This volume alteration of the control spheroid may indicate the growth and spontaneous corruption of the HT-29 spheroid over the cultivation time. 
In contrast, the spheroids that were treated with SN-38 showed monotonic reductions in their spheroid volumes over time for all drug concentrations.

The mean LIVs [Fig.\@ \ref{fig:HT29_Quant}(b)] increased from TD-1 to TD-3 for all drug concentrations, and then decreased from TD-3 to TD-6 for the control sample (P = 0.006, Mann-Whitney test; supplementary Table S2).

The mean \OCDSl in the 1-\uM and 10-\uM SN-38 concentration cases demonstrated significant reductions over time, as illustrated in [Fig.\@ \ref{fig:HT29_Quant}(c)].
The significant reduction in the 1-\uM case was found from TD-3 to TD-6 (P = 0.006, Mann-Whitney test), while that in the 10-\uM was found from TD-1 to TD-3 (P = 0.018). 
For all other SN-38 concentrations, the treatment-time dependency was not observed.
All statistical test results are presented in supplementary Table S2.

The LIV cut-off-based necrotic cell ratio demonstrated a reduction in the necrotic cell ratio from TD-1 to TD-3 for all SN-38 concentrations, and the ratio then became stable from TD-3 to TD-6.  
The reduction in LIV-based necrotic cell ratio from TD-1 to TD-3 was significant in the 1-\uM and 10-\uM cases of SN-38 (P= 0.047 and 0.010, respectively; Mann-Whitney test).

On the other hand, the necrotic cell ratios based on the \OCDSl cut-off remain stable over time for low concentrations (i.e., the control and 0.1 \uM cases), while that of the 10-\uM case increases from TD-1 to TD-3 (P = 0.006).
The ratio of the 1-\um shows a significant increase from TD-3 to TD-6 (P = 0.006).
The results of this test are presented in  Table S2 (supplementary).

The morphological and dynamic-OCT signal alterations of the HT-29 spheroid treated with SN-38 may be related to the drug mechanism of SN-38, as will be discussed later in Section \ref{sec:Diss_HT29_SN-38}.
The OCT intensity, cross-sectional LIV and \OCDSl results for the same data presented in Fig.\@ \ref{fig:HT-29_SN38_Image} are presented in Figs.\@ S4 and S5 in the supplementary material. 
In addition, the responses of one additional HT-29 spheroid under each treatment condition are presented as supplementary Fig.\@ S6.

\section{Discussion}
\label{sec:Diss}
\subsection{Drug effects of PTX appearing on OCT and dynamic OCT}
\label{sec:Diss_MCF7_Taxol}
Dynamic OCT imaging of the control MCF-7 spheroid [Fig.\@ \ref{fig:MCF-7_Taxol_Image} (first column)] showed circular low dynamic signals (LIV and \OCDSl) surrounded by a high-dynamics shell.
Similar image appearances were observed in our previous studies \cite{El-Sadek_BOE2020, El-Sadek_BOE_2021}.
The low and high dynamics observed at the spheroid core and the periphery are believed to highlight the well-known necrotic core (caused by the lack of both oxygen and nutrients supply) and the viable rim of the tumor spheroids, respectively \cite{Hirschhaeuser_BioTechonl2010, Costa_Biotechnol.Advan.2016}. 
In addition, the increase of low dynamics area over the cultivation time [Fig.\@ \ref{fig:MCF-7_Taxol_Image} (first column)] may indicate an increase in the necrotic area over the long cultivation time.

Similarity between the dynamics patterns of the MCF-7 spheroids treated with PTX for one day and that of the control spheroid was observed in Fig.\@ \ref{fig:MCF-7_Taxol_Image} (first to third rows).
This may indicate either the low efficacy of PTX or the resistance of the tumor cells when the treatment time is only one day. 

In addition to the findings mentioned above, the general tendency of the structural and dynamic OCT observed in this study can be explained as follows using the drug mechanism of PTX.
Microtubules (MTs) are polymers that are formed by polymerization of $\alpha$ and $\beta$ tubulin dimers \cite{Gudimchuk_NatRev_2021}.
This MTs have three important functions in the eukaryotic cells. 
First, MTs are the main constituents of the cytoskeleton, which maintains the cell structure and supports the cell with the mechanical resistance to deformation \cite{Hohmann_cells_2019}.
Second, the MTs are involved in mitosis as the main constituents of the mitotic spindles \cite{Gudimchuk_NatRev_2021, Mcintosh_AnnualRev_2002}. 
Third, the MTs act as platforms for intracellular transport of the vesicles and organelles \cite{Vale_cell_2003, Gudimchuk_NatRev_2021}. 

For the functions described above, the MTs exhibit highly dynamic behavior. 
They exert rapid and stochastic phase transitions from extension (by polymerization) to shrinkage (by de-polymerization) \cite{Gudimchuk_NatRev_2021, Tolic_EBioJ_2008}. 
This dynamic instability allows the MTs tips to search for and bind to intracellular components, e.g., capturing chromosomes during mitosis. \cite{Gudimchuk_NatRev_2021, Tolic_EBioJ_2008}. 

PTX is an MT-stabilizing anti-cancer drug.
This drug prevents MT depolymerization and suppresses the dynamic behavior of the MTs \cite{Xiao_Nat1AcadofSci_2006, Weaver_MolCellBio_2014}. 
Therefore, it leads to cell cycle and mitotic arrest, which ultimately leads to cell death \cite{Xiao_Nat1AcadofSci_2006, Weaver_MolCellBio_2014}. 

The structural OCT findings for the MCF-7 spheroid after treatment with PTX showed spheroid shape corruption and volume reduction over the treatment time. 
This spheroid shape corruption may indicate malfunctioning of the first function of the MTs (i.e., maintaining the cell structure and the resistance to cell deformation) under the application of PTX.  
On the other hand, reduction of the spheroid volume may indicate inhibition of mitotic cell divisions (the second function of the MTs) under the effect of PTX.

On the other hand, the  spheroids that were treated with PTX showed reductions in their LIV and \OCDSl signals over the treatment time.
The LIV and \OCDSl reductions may indicate a reduction of the intracellular motility/transport through the MTs (the third function of MTs) and may also indicate  PTX-induced cell death.

\subsection{Drug effects of SN-38 appeared on OCT and dynamics OCT}
\label{sec:Diss_HT29_SN-38}
The dynamics patterns of the TD-3 HT-29 spheroid under all the SN-38 concentrations, except for the control case [Figs.\@ \ref{fig:HT-29_SN38_Image} and \ref{fig:HT29_SN38_MagnifiedImage}] showed a clear low \OCDSl layer at the periphery and this layer corresponded well with the peripheral dead cells (red fluorescence) observed in the FLUO images. 
Among the three concentrations, the 10-\uM case showed evident changes in the mean LIV and the mean \OCDSl from TD-1 to TD-3 [Fig.\@ \ref{fig:HT29_Quant}(b) and (c)], and also showed the smallest spheroid volume among all concentrations, even at TD-1 [Fig.\@ \ref{fig:HT29_Quant}(a)].
These results indicate that OCT and dynamic OCT imaging can demonstrate higher drug effects of 10-\uM SN-38 than the other lower concentrations.

On TD-6, all the SN-38-treated spheroids at all concentrations (0.1, 1.0 and 10 \uM) showed severe structural corruption.
In comparison, on TD-3, these shape corruptions were not evident for all concentrations.
However, the \OCDSl images showed clear low \OCDSl layers at the periphery that were not found in the control case.  
In addition, spheroid volume analysis [Fig.\@ \ref{fig:HT29_Quant}(a)] showed evidently smaller spheroid volumes, even at TD-3.
These findings suggest that dynamic OCT can assess the more moderate drug effect of SN-38 than that can be visualized using non-dynamic OCT.

The alterations of OCT and dynamic OCT signals and the images of the HT-29 spheroids [Figs.\@  \ref{fig:HT29_Quant} and \ref{fig:HT-29_SN38_Image}, respectively] can be understood more by taking the mechanism of SN-38 into account.
Irinotecan is a DNA topoisomerase I (Topo-I) inhibitor that is used to treat the advanced stages of colorectal cancer and it is also approved for second-line treatment in metastatic colorectal cancer \cite{Wallin_OncolRep_2008,Ozawa_CancerDrugRes_2021}. 
SN-38 is an active metabolite of Irinotecan, which is 100- to 1000-fold more active when compared with the Irinotecan itself \cite{Wallin_OncolRep_2008,Kawato_CancerRes_1991}. 
Because Topo-I is an enzyme involved in DNA transcription and replication, the inhibition of Topo-I caused by SN-38 induces DNA damage and a transient S-phase (the cell cycle in which DNA is replicated) arrest. 
SN-38 application then leads to irreversible breaks in the single-strand-DNA that are subsequently transformed into double-strand-DNA breaks. 
As a result of these double-strand-DNA breaks, a series of different apoptotic-related signaling pathways are activated.
This then results in apoptotic cell death \cite{Wallin_OncolRep_2008,Hsiang_CancerRes_1989}.
The damage of DNA and subsequent DNA replication may suppress cell division, and this can account for the smaller spheroid size at TD-3.
This DNA damage then progresses into apoptosis, and this can lead to the structural corruption observed at TD-6.
The thin low-\OCDSl (red) peripheral layers observed in all SN-38 treated spheroids on TD-3 may indicate the onset of apoptosis at the spheroid-drug interface.
We suspect that the low-\OCDSl may indicate the cell death process because the necrosis at the control spheroid center, which is another type of cell death, also exhibits low \OCDSl.
  
\subsection{Combined interpretation of LIV and \OCDSl}
\label{sec:DifLIV_OCDSl}
\label{sec:combinedInterpretation}
\subsubsection{LIV and \OCDSl highlight different aspects of dynamics}
LIV is a measure of the magnitude of the temporal fluctuations in the dB-scaled OCT intensity and it may be sensitive to the magnitude of the intracellular motility. 
On the other hand, the \OCDSl quantifies the speed of the OCT intensity fluctuations and it may be sensitive to the speed of the intracellular motility, as discussed elsewhere \cite{El-Sadek_BOE2020, El-Sadek_BOE_2021}. 
Because of this fundamental difference between them, the LIV and \OCDSl may highlight different cellular processes.
And this may explain why different image patterns appeared in LIV and \OCDSl.
For example, in the control MCF-7 spheroids [Fig.\@ \ref{fig:MCF-7_Taxol_Image}], \OCDSl shows clearer boundaries between the central and peripheral regions than LIV.
In addition, in some cases, \OCDSl shows tessellated domain structures such as those shown in Fig.\@ \ref{fig:MCF-7_Taxol_MagnifiedImage} that are not visible in the corresponding LIV images.

Although the identification of the cellular process that accounts for these LIV and \OCDSl findings remains an open issue, we conclude that LIV and \OCDSl provide different information and thus are complementary methods.

\subsubsection{Combined interpretation of LIV and \OCDSl of tumor cell necrosis}

\begin{figure}
	\centering
	\includegraphics {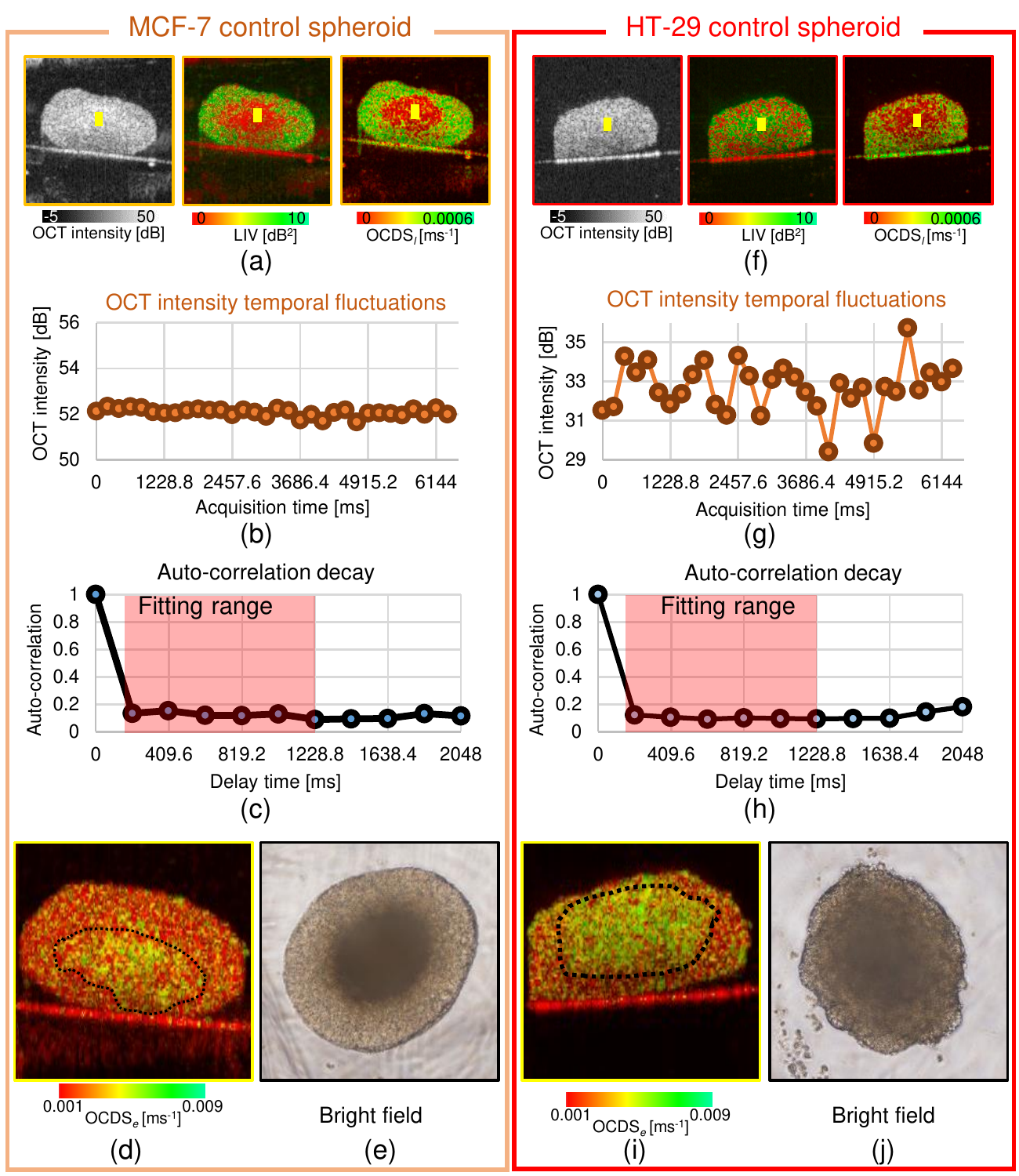}
	\caption{
		Detailed analysis of apparent pattern differences between the control MCF-7 (right) and HT-29 (left) spheroids.
		(a), (f) Cross-sections of the OCT intensity, LIV, and \OCDSl.
		(b), (g) Time course of the OCT intensity in the core areas of the spheroids (as indicated by the yellow boxes).
		MCF-7 shows very low fluctuations, while HT-29 shows high levels of fluctuation. 
		(c), (h) Auto-correlation curves for both spheroids, showing rapid decay as the auto-correlation became very low at the first sampling point of the delay time (204.8 ms).
		(d), (i) \OCDSe cross-sections of the same spheroids, showing the fast decay (green)at the center parts of the spheroid. 
		These \OCDSe images were computed from a cross-sectional frame sequence of 350 frames acquired at a single location.
		(e), (j) Bright-field microscopy images of the spheroids, which show the necrotic cores as dark central region.
	}
	\label{fig:LIV-OCDSl_Diff}	
\end{figure}
By comparing the dynamics OCT images of the TD-1 control spheroids of MCF-7 and HT-29 [Fig.\@ \ref{fig:LIV-OCDSl_Diff}], it can be found that \OCDSl exhibit similar appearances for both cell types.
Namely, the center region shows low \OCDSl, while the periphery exhibits high \OCDSl.
In contrast, the LIV patterns of these spheroids are opposite to each other.
Namely, the MCF-7 spheroid shows low and high LIV at its central and peripheral regions, respectively, while the HT-29 spheroid showed high and low LIV, respectively, in the corresponding regions  [Fig.\@ \ref{fig:LIV-OCDSl_Diff} (a) and (f)].

To further investigate this difference, the time sequence of the OCT intensity [Fig.\@ \ref{fig:LIV-OCDSl_Diff}(b) and (g)] and the autocorrelation curves [Fig.\@ \ref{fig:LIV-OCDSl_Diff}(c) and (h)] are presented; these figures show the averaged intensity and the autocorrelation curves in the yellow boxed regions indicated in the OCT and dynamic OCT images.
Both autocorrelation curves show that the correlation is strongly decayed even at the first computing point (204.8 ms), and thus the signal fluctuations in both cases are very fast.
In comparison, the time sequence OCT intensities show that the signal fluctuation magnitude is small for MCF-7 but it is large for the HT-29 spheroid.

To provide further validation of this fast autocorrelation decay, a single B-scan location in the same spheroids was scanned 350 times in 4.48 s, i.e., it was scanned with much higher temporal density.
This is the protocol that we presented for our original 2D dynamic imaging, and it enables yet another dynamic OCT contrast, the fast/early correlation decay speed (\OCDSe) \cite{El-Sadek_BOE2020}.
This contrast is similar to \OCDSl, but the slope of the correlation decay is defined at an earlier time range [12.8, 64 ms] than \OCDSl, and thus it is more directly sensitive to the fast signal fluctuations.
The \OCDSe images of the spheroids are presented in Fig.\@ \ref{fig:LIV-OCDSl_Diff} (d) and (i).
As shown in these images, the \OCDSe is high at the center regions of both spheroids.
This represents further proof that the dynamics at the center regions are fast.

It is known that the center part of the spheroid becomes necrotic because of the lack of oxygen and other nutrients \cite{Sutherland_CancerRes1986, Hirschhaeuser_BioTechonl2010, Vinci_BMCBiol2012,Lee_Biosensors_2021}.
In addition, it has been reported previously that cell necrosis shows a fast correlation decay \cite{Farhat2011SPIE}.
As a result, we suspect that the center regions of both spheroids are in the necrotic process.

It is also known that the necrotic process takes a long time after the cell death starts \cite{Majno_Am_J_Pathol_1995}.
The difference in the magnitude of the signal fluctuations may be accounted for by the time after the onset of cell death.
Namely, we can assume that the onset of cell death for this particular MCF-7 spheroid was early, and hence the necrotic process has been progressed, this results in the low fluctuation magnitude.
One the other hand, the cell death onset of the HT-29 spheroid may be late, which means that the signal fluctuation magnitude is still large.

To further investigate these differences in the spheroid core activities, bright field microscopic images acquired using (IX71, Olympus) microscope with an objective with 4x magnification and an NA of 0.13 from the same samples that were presented in Fig. \ref{fig:LIV-OCDSl_Diff} are shown in Fig.\@ \ref{fig:LIV-OCDSl_Diff}. 

The bright field image of the MCF-7 spheroid [Fig.\@ \ref{fig:LIV-OCDSl_Diff} (e)] shows a dark appearance at its center, which is a so-called necrotic core, with very clear boundaries surrounded by a bright appearance (viable cells). 
In contrast, the HT-29 spheroid [Fig.\@ \ref{fig:LIV-OCDSl_Diff} (j)] shows a diffusive border.
This may indicate that the necrotic core has not completely formed yet.
These bright field images thus support our hypothesis.

In addition, our previous study involving time lapse imaging of the MCF-7 spheroid demonstrated that the LIV at the central region was high at the early time points and it gradually degraded along hours \cite{El-Sadek_BOE2020}. 
These results may provide further support for our hypothesis.

\subsection{Other dynamic OCT methods}
Including our LIV and \OCDSl, there are several dynamic OCT methods.
These methods can be classified into two categories: magnitude evaluations and speed evaluations of the OCT signal fluctuations.
The latter category can be categorized further into two subcategories that include autocorrelation analysis and time spectrum analysis.

The magnitude analysis may include following research works.
The cumulative sum (cumsum) method demonstrated by Scholler quantifies the signal fluctuation magnitude using a Brownian bridge model, and it has high tolerance to bulk sample motion \cite{Scholler2019OpEx}.
The motility amplitude method demonstrated by Oldenberg is also a metric to quantify the signal fluctuation magnitude \cite{Oldenburg_Optica2015}.
These modalities are expected to visualize the fluctuation magnitudes of tissue and cellular dynamics.

The correlation analyses can quantify the speeds of the tissue dynamics.
Leroux \etal computed the autocorrelation decay of an OCT signal and fitted it with a biexponential function, so that cellular dynamics with multiple speed were quantified \cite{Leroux_BOE2016}.

Spectrum analysis is another method that can be used to quantify the temporal characteristics of the OCT signal fluctuations and tissue dynamics.
The spectroscopic analysis demonstrated by Oldenburg \etal is one of the earliest demonstrations of this approach \cite{Oldenburg_Optica2015}.
Apelian \etal combined a signal frequency analysis with a high-resolution full-field OCT imaging, and enabled high-resolution and functional imaging of \exvivo animal tissues \cite{Apelian2016BOE}. 
Scholler et al. visualized the activity of retinal organoids through power-spectrum analysis of time-sequential OCT signals \cite {Scholler2020LSA}.
Recently, M\"unter \etal used a time-frequency analysis to enable three-dimensional visualization of tissue dynamics, and visualized the cellular scale dynamics of \exvivo animal tissues \cite{Munter2020OL, Munter2021BOE}.
Both the correlation and spectral analysis methods may contrast the speed of the tissue dynamics.

In the present study, we used two methods: LIV and \OCDSl. 
LIV is a method to quantify the magnitude of the signal fluctuation, quantifies the speed of the fluctuation.
As discussed in Section \ref{sec:combinedInterpretation}, the combination of these two methods provides a comprehensive interpretation of the dynamic processes occurring in the tissue.

\section{Conclusion}
The spatial patterns of the anti-cancer drug responses of two types of human tumor spheroid, including those derived from breast (MCF-7) and colon (HT-29) cancer cell-lines have been visualized using two types of dynamic OCT algorithms.
The response patterns observed in dynamic OCT images were consistent with the corresponding fluorescence microscopy patterns.    
In addition, the spheroid volume and mean dynamic OCT signal values were computed.
The time-course changes in these values revealed different trends among the two types of spheroid and the anti-cancer drugs.

In conclusion, dynamic OCT can be used to highlight the difference in drug response patterns among different tumor spheroids and different drug types.
Therefore, dynamic OCT can be a useful tool for anti-cancer drug testing and for the optimal selection of anti-cancer drugs.

\section*{Acknowledgments}
The authors greatly appreciate the fruitful technical discussions that were held with Hiroyuki Sangu (Yokogawa), Arata Miyazawa, Atsushi Kubota and Renzo Ikeda (Sky technology), Akihiro Shitoh and Yuichi Inoue (Optosigma), Masato Takaya (Tatsuta), and Naoki Fukutake (Nikon).

\section*{Funding}
Core Research for Evolutional Science and Technology (JPMJCR2105); Japan Society for the Promotion of Science (18H01893, 21H01836, 22K04962); Austrian Science Fund (Schr\"odinger grand J4460); Japan Science and Technology Agency (JPMJMI18G8).

\section*{Disclosures}
Abd El-Sadek,  Makita, Mukherjee, Yasuno: Yokogawa Electric Corp. (F), Sky Technology (F), Nikon (F), Kao Corp. (F), Topcon (F).
Shen, Mori, Lichtenegger, Matsusaka: None.

\section*{Data availability} Data underlying the results presented in this paper are not publicly available at this time but may be obtained from the authors upon reasonable request.

\section*{Supplemental document}
Six supplementary figures (Figs.\@ S1 to S6) and two supplementary tables (Tables S1 and S2) are available in Supplement 1.

\bibliography{References}

\begin{thebibliography}{10}
\newcommand{\enquote}[1]{``#1''}

\bibitem{Sungung2020CancerJClin}
H.~Sung, J.~Ferlay, R.~L. Siegel, M.~Laversanne, I.~Soerjomataram, A.~Jemal,
  and F.~Bray, \enquote{Global {Cancer} {Statistics} 2020: {GLOBOCAN}
  {Estimates} of {Incidence} and {Mortality} {Worldwide} for 36 {Cancers} in
  185 {Countries},} {\protect\JournalTitle{CA: A Cancer Journal for
  Clinicians}} \textbf{71}, 209--249 (2021). \_eprint:
  https://onlinelibrary.wiley.com/doi/pdf/10.3322/caac.21660.

\bibitem{Bray2021Cancer}
F.~Bray, M.~Laversanne, E.~Weiderpass, and I.~Soerjomataram, \enquote{The
  ever-increasing importance of cancer as a leading cause of premature death
  worldwide,} {\protect\JournalTitle{Cancer}} \textbf{127}, 3029--3030 (2021).
  \_eprint: https://onlinelibrary.wiley.com/doi/pdf/10.1002/cncr.33587.

\bibitem{Costa_Biotechnol.Advan.2016}
E.~C. Costa, A.~F. Moreira, D.~de~Melo-Diogo, V.~M. Gaspar, M.~P. Carvalho, and
  I.~J. Correia, \enquote{{3D} tumor spheroids: an overview on the tools and
  techniques used for their analysis,} {\protect\JournalTitle{Biotechnology
  Advances}} \textbf{34}, 1427--1441 (2016).

\bibitem{Shahi_AssayDrugDevTechnol2019}
P.~Shahi~Thakuri, M.~Gupta, M.~Plaster, and H.~Tavana, \enquote{Quantitative
  {Size}-{Based} {Analysis} of {Tumor} {Spheroids} and {Responses} to
  {Therapeutics},} {\protect\JournalTitle{Assay and Drug Development
  Technologies}} \textbf{17}, 140--149 (2019).

\bibitem{Han_CancerCellInt2021}
S.~J. Han, S.~Kwon, and K.~S. Kim, \enquote{Challenges of applying
  multicellular tumor spheroids in preclinical phase,}
  {\protect\JournalTitle{Cancer Cell International}} \textbf{21}, 152 (2021).

\bibitem{Lee_Biosensors_2021}
K.-H. Lee and T.-H. Kim, \enquote{Recent {Advances} in {Multicellular} {Tumor}
  {Spheroid} {Generation} for {Drug} {Screening},}
  {\protect\JournalTitle{Biosensors}} \textbf{11}, 445 (2021). Number: 11
  Publisher: Multidisciplinary Digital Publishing Institute.

\bibitem{kobayashi_PNAS1993}
H.~Kobayashi, S.~Man, C.~H. Graham, S.~J. Kapitain, B.~A. Teicher, and R.~S.
  Kerbel, \enquote{Acquired multicellular-mediated resistance to alkylating
  agents in cancer.} {\protect\JournalTitle{Proceedings of the National Academy
  of Sciences}} \textbf{90}, 3294--3298 (1993). Publisher: National Academy of
  Sciences Section: Research Article.

\bibitem{Ivascu_BiomolScreen2006}
A.~Ivascu and M.~Kubbies, \enquote{Rapid {Generation} of {Single}-{Tumor}
  {Spheroids} for {High}-{Throughput} {Cell} {Function} and {Toxicity}
  {Analysis},} {\protect\JournalTitle{Journal of Biomolecular Screening}}
  \textbf{11}, 922--932 (2006). Publisher: SAGE Publications Inc STM.

\bibitem{Dubois_Oncotarget2017}
C.~Dubois, R.~Dufour, P.~Daumar, C.~Aubel, C.~Szczepaniak, C.~Blavignac,
  E.~Mounetou, F.~Penault-Llorca, and M.~Bamdad, \enquote{Development and
  cytotoxic response of two proliferative {MDA}-{MB}-231 and non-proliferative
  {SUM1315} three-dimensional cell culture models of triple-negative basal-like
  breast cancer cell lines,} {\protect\JournalTitle{Oncotarget}} \textbf{8},
  95316--95331 (2017).

\bibitem{Thakuri_IEEE2016}
P.~S. Thakuri, S.~L. Ham, and H.~Tavana, \enquote{Microprinted tumor spheroids
  enable anti-cancer drug screening,} {\protect\JournalTitle{Annual
  International Conference of the IEEE Engineering in Medicine and Biology
  Society. IEEE Engineering in Medicine and Biology Society. Annual
  International Conference}} \textbf{2016}, 4177--4180 (2016).

\bibitem{Degrandis_FrontOncol2021}
R.~A. De~Grandis, K.~M. Oliveira, A.~P.~M. Guedes, P.~W.~S. dos Santos, A.~F.
  Aissa, A.~A. Batista, and F.~R. Pavan, \enquote{A {Novel} {Ruthenium}({II})
  {Complex} {With} {Lapachol} {Induces} {G2}/{M} {Phase} {Arrest} {Through}
  {Aurora}-{B} {Kinase} {Down}-{Regulation} and {ROS}-{Mediated}
  {Apoptosis} in {Human} {Prostate} {Adenocarcinoma} {Cells},}
  {\protect\JournalTitle{Frontiers in Oncology}} \textbf{11}, 2148 (2021).

\bibitem{Jep2017POlSONE}
M.~Jeppesen, G.~Hagel, A.~Glenthoj, B.~Vainer, P.~Ibsen, H.~Harling,
  O.~Thastrup, L.~N. Jørgensen, and J.~Thastrup, \enquote{Short-term spheroid
  culture of primary colorectal cancer cells as an in vitro model for
  personalizing cancer medicine,} {\protect\JournalTitle{PLOS ONE}}
  \textbf{12}, e0183074 (2017). Publisher: Public Library of Science.

\bibitem{plummer2019SciRep}
S.~Plummer, S.~Wallace, G.~Ball, R.~Lloyd, P.~Schiapparelli,
  A.~Quiñones-Hinojosa, T.~Hartung, and D.~Pamies, \enquote{A {Human}
  {iPSC}-derived {3D} platform using primary brain cancer cells to study drug
  development and personalized medicine,} {\protect\JournalTitle{Scientific
  Reports}} \textbf{9}, 1407 (2019). Number: 1 Publisher: Nature Publishing
  Group.

\bibitem{pampaloni2013Cell}
F.~Pampaloni, N.~Ansari, and E.~H.~K. Stelzer, \enquote{High-resolution deep
  imaging of live cellular spheroids with light-sheet-based fluorescence
  microscopy,} {\protect\JournalTitle{Cell and Tissue Research}} \textbf{352},
  161--177 (2013).

\bibitem{Mittler2017FRONTONCOL}
F.~Mittler, P.~Obeïd, A.~V. Rulina, V.~Haguet, X.~Gidrol, and M.~Y. Balakirev,
  \enquote{High-{Content} {Monitoring} of {Drug} {Effects} in a {3D} {Spheroid}
  {Model},} {\protect\JournalTitle{Frontiers in Oncology}} \textbf{7} (2017).
  Publisher: Frontiers.

\bibitem{Yang2019MaterDes}
W.~Yang, S.~Cai, Z.~Yuan, Y.~Lai, H.~Yu, Y.~Wang, and L.~Liu,
  \enquote{Mask-free generation of multicellular {3D} heterospheroids array for
  high-throughput combinatorial anti-cancer drug screening,}
  {\protect\JournalTitle{Materials \& Design}} \textbf{183}, 108182 (2019).

\bibitem{Baek2016OncoTargetsTher}
N.~Baek, O.~W. Seo, M.~Kim, J.~Hulme, and S.~S.~A. An, \enquote{Monitoring the
  effects of doxorubicin on {3D}-spheroid tumor cells in real-time,}  (2016).

\bibitem{Zoetemelk_SciRep2019}
M.~Zoetemelk, M.~Rausch, D.~J. Colin, O.~Dormond, and P.~Nowak-Sliwinska,
  \enquote{Short-term {3D} culture systems of various complexity for treatment
  optimization of colorectal carcinoma,} {\protect\JournalTitle{Scientific
  Reports}} \textbf{9}, 7103 (2019). Number: 1 Publisher: Nature Publishing
  Group.

\bibitem{Drexler2015SprinInt}
W.~Drexler and J.~G. Fujimoto, eds., \emph{Optical {Coherence} {Tomography}:
  {Technology} and {Applications}} (Springer International Publishing, 2015),
  2nd ed.

\bibitem{Aguirre2015OL}
A.~D. Aguirre, C.~Zhou, H.-C. Lee, O.~O. Ahsen, and J.~G. Fujimoto,
  \enquote{Optical {Coherence} {Microscopy},} in \emph{Optical {Coherence}
  {Tomography}: {Technology} and {Applications},}  W.~Drexler and J.~G.
  Fujimoto, eds. (Springer International Publishing, Cham, 2015), pp. 865--911.

\bibitem{Vermeer2014BOE}
K.~A. Vermeer, J.~Mo, J.~J.~A. Weda, H.~G. Lemij, and J.~F.~d. Boer,
  \enquote{Depth-resolved model-based reconstruction of attenuation
  coefficients in optical coherence tomography,}
  {\protect\JournalTitle{Biomedical Optics Express}} \textbf{5}, 322--337
  (2014). Publisher: Optical Society of America.

\bibitem{Gong2020JBO}
P.~Gong, M.~Almasian, G.~van Soest, D.~M. de~Bruin, T.~G. van Leeuwen, D.~D.
  Sampson, and D.~J. Faber, \enquote{Parametric imaging of attenuation by
  optical coherence tomography: review of models, methods, and clinical
  translation,} {\protect\JournalTitle{Journal of Biomedical Optics}}
  \textbf{25} (2020).

\bibitem{VanDerMeer_2010}
F.~J. van~der Meer, D.~J. Faber, M.~C.~G. Aalders, A.~A. Poot, I.~Vermes, and
  T.~G. van Leeuwen, \enquote{Apoptosis- and necrosis-induced changes in light
  attenuation measured by optical coherence tomography,}
  {\protect\JournalTitle{Lasers in Medical Science}} \textbf{25}, 259--267
  (2010).

\bibitem{DeBoer2017BOE}
J.~F. de~Boer, C.~K. Hitzenberger, and Y.~Yasuno, \enquote{Polarization
  sensitive optical coherence tomography – a review [{Invited}],}
  {\protect\JournalTitle{Biomed. Opt. Express}} \textbf{8}, 1838--1873 (2017).

\bibitem{Sugiyama2015BOE}
S.~Sugiyama, Y.-J. Hong, D.~Kasaragod, S.~Makita, S.~Uematsu, Y.~Ikuno,
  M.~Miura, and Y.~Yasuno, \enquote{Birefringence imaging of posterior eye by
  multi-functional {Jones} matrix optical coherence tomography,}
  {\protect\JournalTitle{Biomedical Optics Express}} \textbf{6}, 4951--4974
  (2015). Publisher: Optical Society of America.

\bibitem{Li2017BOE}
E.~Li, S.~Makita, Y.-J. Hong, D.~Kasaragod, and Y.~Yasuno,
  \enquote{Three-dimensional multi-contrast imaging of in vivo human skin by
  {Jones} matrix optical coherence tomography,} {\protect\JournalTitle{Biomed.
  Opt. Express}} \textbf{8}, 1290--1305 (2017).

\bibitem{Gotzinger2008OE}
E.~Götzinger, M.~Pircher, W.~Geitzenauer, C.~Ahlers, B.~Baumann, S.~Michels,
  U.~Schmidt-Erfurth, and C.~K. Hitzenberger, \enquote{Retinal pigment
  epithelium segmentation by polarization sensitive optical coherence
  tomography,} {\protect\JournalTitle{Optics Express}} \textbf{16},
  16410--16422 (2008). Publisher: Optical Society of America.

\bibitem{Makita2014OL}
S.~Makita, Y.-J. Hong, M.~Miura, and Y.~Yasuno, \enquote{Degree of polarization
  uniformity with high noise immunity using polarization-sensitive optical
  coherence tomography,} {\protect\JournalTitle{Opt. Lett.}} \textbf{39},
  6783--6786 (2014).

\bibitem{Miura_SciRep2021}
M.~Miura, S.~Makita, Y.~Yasuno, T.~Iwasaki, S.~Azuma, T.~Mino, and
  T.~Yamaguchi, \enquote{Evaluation of retinal pigment epithelium changes in
  serous pigment epithelial detachment in age-related macular degeneration,}
  {\protect\JournalTitle{Scientific Reports}} \textbf{11}, 2764 (2021).
  Bandiera\_abtest: a Cc\_license\_type: cc\_by Cg\_type: Nature Research
  Journals Number: 1 Primary\_atype: Research Publisher: Nature Publishing
  Group Subject\_term: Macular degeneration;Translational research
  Subject\_term\_id: macular-degeneration;translational-research.

\bibitem{Kennedy2014BOE}
B.~F. Kennedy, R.~A. McLaughlin, K.~M. Kennedy, L.~Chin, A.~Curatolo, A.~Tien,
  B.~Latham, C.~M. Saunders, and D.~D. Sampson, \enquote{Optical coherence
  micro-elastography: mechanical-contrast imaging of tissue microstructure,}
  {\protect\JournalTitle{Biomed. Opt. Express}} \textbf{5}, 2113--2124 (2014).

\bibitem{kennedy_SciRep2015}
K.~M. Kennedy, L.~Chin, R.~A. McLaughlin, B.~Latham, C.~M. Saunders, D.~D.
  Sampson, and B.~F. Kennedy, \enquote{Quantitative micro-elastography: imaging
  of tissue elasticity using compression optical coherence elastography,}
  {\protect\JournalTitle{Scientific Reports}} \textbf{5}, 15538 (2015).
  Bandiera\_abtest: a Cc\_license\_type: cc\_by Cg\_type: Nature Research
  Journals Number: 1 Primary\_atype: Research Publisher: Nature Publishing
  Group Subject\_term: Biomedical engineering;Biophotonics;Breast cancer
  Subject\_term\_id: biomedical-engineering;biophotonics;breast-cancer.

\bibitem{Gubarkova_BOE2019}
E.~V. Gubarkova, A.~A. Sovetsky, V.~Y. Zaitsev, A.~L. Matveyev, D.~A.
  Vorontsov, M.~A. Sirotkina, L.~A. Matveev, A.~A. Plekhanov, N.~P. Pavlova,
  S.~S. Kuznetsov, A.~Y. Vorontsov, E.~V. Zagaynova, and N.~D. Gladkova,
  \enquote{{OCT}-elastography-based optical biopsy for breast cancer
  delineation and express assessment of morphological/molecular subtypes,}
  {\protect\JournalTitle{Biomedical Optics Express}} \textbf{10}, 2244--2263
  (2019).

\bibitem{Miyazawa2019BOE}
A.~Miyazawa, S.~Makita, E.~Li, K.~Yamazaki, M.~Kobayashi, S.~Sakai, and
  Y.~Yasuno, \enquote{Polarization-sensitive optical coherence elastography,}
  {\protect\JournalTitle{Biomed. Opt. Express}} \textbf{10}, 5162--5181 (2019).

\bibitem{Apelian2016BOE}
C.~Apelian, F.~Harms, O.~Thouvenin, and A.~C. Boccara, \enquote{Dynamic full
  field optical coherence tomography: subcellular metabolic contrast revealed
  in tissues by interferometric signals temporal analysis,}
  {\protect\JournalTitle{Biomed. Opt. Express}} \textbf{7}, 1511--1524 (2016).

\bibitem{Thouvenin2017ApplSci}
O.~Thouvenin, C.~Apelian, A.~Nahas, M.~Fink, and C.~Boccara,
  \enquote{Full-{Field} {Optical} {Coherence} {Tomography} as a {Diagnosis}
  {Tool}: {Recent} {Progress} with {Multimodal} {Imaging},}
  {\protect\JournalTitle{Appl. Sci.}} \textbf{7}, 236 (2017).

\bibitem{Scholler2020LSA}
J.~Scholler, K.~Groux, O.~Goureau, J.-A. Sahel, M.~Fink, S.~Reichman,
  C.~Boccara, and K.~Grieve, \enquote{Dynamic full-field optical coherence
  tomography: {3D} live-imaging of retinal organoids,}
  {\protect\JournalTitle{Light: Science \& Applications}} \textbf{9}, 140
  (2020). Bandiera\_abtest: a Cc\_license\_type: cc\_by Cg\_type: Nature
  Research Journals Number: 1 Primary\_atype: Research Publisher: Nature
  Publishing Group Subject\_term: Imaging and sensing;Interference
  microscopy;Wide-field fluorescence microscopy Subject\_term\_id:
  imaging-and-sensing;interference-microscopy;wide-field-fluorescence-microscopy.

\bibitem{leung2020BOE}
H.~M. Leung, M.~L. Wang, H.~Osman, E.~Abouei, C.~MacAulay, M.~Follen, J.~A.
  Gardecki, and G.~J. Tearney, \enquote{Imaging intracellular motion with
  dynamic micro-optical coherence tomography,} {\protect\JournalTitle{Biomed.
  Opt. Express}} \textbf{11}, 2768--2778 (2020).

\bibitem{El-Sadek_BOE2020}
I.~A. El-Sadek, A.~Miyazawa, L.~T.-W. Shen, S.~Makita, S.~Fukuda, S.~Fukuda,
  T.~Yamashita, Y.~Oka, P.~Mukherjee, S.~Matsusaka, T.~Oshika, H.~Kano, and
  Y.~Yasuno, \enquote{Optical coherence tomography-based tissue dynamics
  imaging for longitudinal and drug response evaluation of tumor spheroids,}
  {\protect\JournalTitle{Biomedical Optics Express}} \textbf{11}, 6231--6248
  (2020). Publisher: Optical Society of America.

\bibitem{Scholler_LightSciAppl_2020}
J.~Scholler, K.~Groux, O.~Goureau, J.-A. Sahel, M.~Fink, S.~Reichman,
  C.~Boccara, and K.~Grieve, \enquote{Dynamic full-field optical coherence
  tomography: {3D} live-imaging of retinal organoids,}
  {\protect\JournalTitle{Light: Science \& Applications}} \textbf{9}, 140
  (2020). Bandiera\_abtest: a Cc\_license\_type: cc\_by Cg\_type: Nature
  Research Journals Number: 1 Primary\_atype: Research Publisher: Nature
  Publishing Group Subject\_term: Imaging and sensing;Interference
  microscopy;Wide-field fluorescence microscopy Subject\_term\_id:
  imaging-and-sensing;interference-microscopy;wide-field-fluorescence-microscopy.

\bibitem{Munter2020OL}
M.~Münter, M.~v. Endt, M.~Pieper, M.~Pieper, M.~Casper, M.~Ahrens, M.~Ahrens,
  T.~Kohlfaerber, R.~Rahmanzadeh, P.~König, P.~König, G.~Hüttmann,
  G.~Hüttmann, G.~Hüttmann, H.~Schulz-Hildebrandt, H.~Schulz-Hildebrandt, and
  H.~Schulz-Hildebrandt, \enquote{Dynamic contrast in scanning microscopic
  {OCT},} {\protect\JournalTitle{Optics Letters}} \textbf{45}, 4766--4769
  (2020). Publisher: Optical Society of America.

\bibitem{Kurokawa2020Neuro}
K.~Kurokawa, J.~A. Crowell, F.~Zhang, and D.~T. Miller, \enquote{Suite of
  methods for assessing inner retinal temporal dynamics across spatial and
  temporal scales in the living human eye,}
  {\protect\JournalTitle{Neurophotonics}} \textbf{7}, 015013 (2020). Publisher:
  International Society for Optics and Photonics.

\bibitem{Munter2021BOE}
M.~Münter, M.~Münter, M.~Pieper, M.~Pieper, T.~Kohlfaerber, E.~Bodenstorfer,
  M.~Ahrens, M.~Ahrens, C.~Winter, R.~Huber, P.~König, P.~König,
  G.~Hüttmann, G.~Hüttmann, G.~Hüttmann, H.~Schulz-Hildebrandt,
  H.~Schulz-Hildebrandt, and H.~Schulz-Hildebrandt, \enquote{Microscopic
  optical coherence tomography ({mOCT}) at 600 {kHz} for {4D} volumetric
  imaging and dynamic contrast,} {\protect\JournalTitle{Biomedical Optics
  Express}} \textbf{12}, 6024--6039 (2021). Publisher: Optica Publishing Group.

\bibitem{ElSadek_SPIE2021}
I.~G.~A. El-Sadek, A.~Miyazawa, L.~T.~W. Shen, S.~Makita, P.~Mukherjee,
  S.~Matsusaka, and Y.~Yasuno, \enquote{{OCT} based cross-sectional and
  three-dimensional dynamics imaging for visualization and quantification of
  tumor spheroid activity,} in \emph{Optical {Coherence} {Tomography} and
  {Coherence} {Domain} {Optical} {Methods} in {Biomedicine} {XXV},}  vol. 11630
  (International Society for Optics and Photonics, 2021), p. 116301E.

\bibitem{El-Sadek_BOE_2021}
I.~A. El-Sadek, I.~A. El-Sadek, A.~Miyazawa, L.~T.-W. Shen, S.~Makita,
  P.~Mukherjee, A.~Lichtenegger, A.~Lichtenegger, S.~Matsusaka, and Y.~Yasuno,
  \enquote{Three-dimensional dynamics optical coherence tomography for tumor
  spheroid evaluation,} {\protect\JournalTitle{Biomedical Optics Express}}
  \textbf{12}, 6844--6863 (2021). Publisher: Optical Society of America.

\bibitem{Newschaffer_Lancet2001}
C.~J. Newschaffer, A.~Topham, T.~Herzberg, S.~Weiner, and D.~S. Weinberg,
  \enquote{Risk of colorectal cancer after breast cancer,}
  {\protect\JournalTitle{The Lancet}} \textbf{357}, 837--840 (2001).

\bibitem{Rodriguez_BiomedCompMeth2013}
J.~Rodriguez-Soler, J.~M. Sabater-Navarro, N.~García, and F.~J. Amoros,
  \enquote{Ultrasound based application for intraglandular mapping of breast
  cancer,} {\protect\JournalTitle{Computer Methods and Programs in
  Biomedicine}} \textbf{112}, 293--301 (2013).

\bibitem{Vakili_HematolOncolStemCellRes2014}
S.~Vakili, M.~Sharbatdaran, A.~Noorbaran, S.~Siadati, D.~Moslemi, and
  S.~Shafahi, \enquote{A {Case} of {Colon} {Cancer} with {Breast} {Metastasis}
  and {Krukenberg} {Tumor},} {\protect\JournalTitle{International Journal of
  Hematology-Oncology and Stem Cell Research}} \textbf{8}, 46--50 (2014).

\bibitem{Majid_BiomedCOMPUTMETH2014}
A.~Majid, S.~Ali, M.~Iqbal, and N.~Kausar, \enquote{Prediction of human breast
  and colon cancers from imbalanced data using nearest neighbor and support
  vector machines,} {\protect\JournalTitle{Computer Methods and Programs in
  Biomedicine}} \textbf{113}, 792--808 (2014).

\bibitem{Hirschhaeuser_BioTechonl2010}
F.~Hirschhaeuser, H.~Menne, C.~Dittfeld, J.~West, W.~Mueller-Klieser, and L.~A.
  Kunz-Schughart, \enquote{Multicellular tumor spheroids: {An} underestimated
  tool is catching up again,} {\protect\JournalTitle{Journal of Biotechnology}}
  \textbf{148}, 3--15 (2010).

\bibitem{Gudimchuk_NatRev_2021}
N.~B. Gudimchuk and J.~R. McIntosh, \enquote{Regulation of microtubule
  dynamics, mechanics and function through the growing tip,}
  {\protect\JournalTitle{Nature Reviews Molecular Cell Biology}} \textbf{22},
  777--795 (2021). Number: 12 Publisher: Nature Publishing Group.

\bibitem{Hohmann_cells_2019}
T.~Hohmann and F.~Dehghani, \enquote{The {Cytoskeleton}—{A} {Complex}
  {Interacting} {Meshwork},} {\protect\JournalTitle{Cells}} \textbf{8}, 362
  (2019).

\bibitem{Mcintosh_AnnualRev_2002}
J.~R. McIntosh, E.~L. Grishchuk, and R.~R. West,
  \enquote{Chromosome-microtubule interactions during mitosis,}
  {\protect\JournalTitle{Annual Review of Cell and Developmental Biology}}
  \textbf{18}, 193--219 (2002).

\bibitem{Vale_cell_2003}
R.~Vale, \enquote{The {Molecular} {Motor} {Toolbox} for {Intracellular}
  {Transport},} {\protect\JournalTitle{Cell}} \textbf{112}, 467--480 (2003).

\bibitem{Tolic_EBioJ_2008}
I.~M. Tolić-Nørrelykke, \enquote{Push-me-pull-you: how microtubules organize
  the cell interior,} {\protect\JournalTitle{European Biophysics Journal}}
  \textbf{37}, 1271--1278 (2008).

\bibitem{Xiao_Nat1AcadofSci_2006}
H.~Xiao, P.~Verdier-Pinard, N.~Fernandez-Fuentes, B.~Burd, R.~Angeletti,
  A.~Fiser, S.~B. Horwitz, and G.~A. Orr, \enquote{Insights into the mechanism
  of microtubule stabilization by {Taxol},} {\protect\JournalTitle{Proceedings
  of the National Academy of Sciences}} \textbf{103}, 10166--10173 (2006).
  Publisher: National Academy of Sciences Section: Biological Sciences.

\bibitem{Weaver_MolCellBio_2014}
B.~A. Weaver, \enquote{How {Taxol}/paclitaxel kills cancer cells,}
  {\protect\JournalTitle{Molecular Biology of the Cell}} \textbf{25},
  2677--2681 (2014). Publisher: American Society for Cell Biology (mboc).

\bibitem{Wallin_OncolRep_2008}
A.~Wallin, J.~Svanvik, B.~Holmlund, L.~Ferreud, and X.-F. Sun,
  \enquote{Anticancer effect of {SN}-38 on colon cancer cell lines with
  different metastatic potential,} {\protect\JournalTitle{Oncology Reports}}
  \textbf{19}, 1493--1498 (2008). Publisher: Spandidos Publications.

\bibitem{Ozawa_CancerDrugRes_2021}
S.~Ozawa, T.~Miura, J.~Terashima, and W.~Habano, \enquote{Cellular irinotecan
  resistance in colorectal cancer and overcoming irinotecan refractoriness
  through various combination trials including {DNA} methyltransferase
  inhibitors: a review,} {\protect\JournalTitle{Cancer Drug Resistance}}
  \textbf{4}, 946--964 (2021). Publisher: OAE Publishing Inc.

\bibitem{Kawato_CancerRes_1991}
Y.~Kawato, M.~Aonuma, Y.~Hirota, H.~Kuga, and K.~Sato, \enquote{Intracellular
  roles of {SN}-38, a metabolite of the camptothecin derivative {CPT}-11, in
  the antitumor effect of {CPT}-11,} {\protect\JournalTitle{Cancer Research}}
  \textbf{51}, 4187--4191 (1991).

\bibitem{Hsiang_CancerRes_1989}
Y.~H. Hsiang, M.~G. Lihou, and L.~F. Liu, \enquote{Arrest of replication forks
  by drug-stabilized topoisomerase {I}-{DNA} cleavable complexes as a mechanism
  of cell killing by camptothecin,} {\protect\JournalTitle{Cancer Research}}
  \textbf{49}, 5077--5082 (1989).

\bibitem{Sutherland_CancerRes1986}
R.~M. Sutherland, B.~Sordat, J.~Bamat, H.~Gabbert, B.~Bourrat, and
  W.~Mueller-Klieser, \enquote{Oxygenation and differentiation in multicellular
  spheroids of human colon carcinoma,} {\protect\JournalTitle{Cancer Research}}
  \textbf{46}, 5320--5329 (1986).

\bibitem{Vinci_BMCBiol2012}
M.~Vinci, S.~Gowan, F.~Boxall, L.~Patterson, M.~Zimmermann, W.~Court, C.~Lomas,
  M.~Mendiola, D.~Hardisson, and S.~A. Eccles, \enquote{Advances in
  establishment and analysis of three-dimensional tumor spheroid-based
  functional assays for target validation and drug evaluation,}
  {\protect\JournalTitle{BMC Biology}} \textbf{10}, 29 (2012).

\bibitem{Farhat2011SPIE}
G.~Farhat, A.~Mariampillai, V.~X.~D. Yang, G.~J. Czarnota, and M.~C. Kolios,
  \enquote{Optical coherence tomography speckle decorrelation for detecting
  cell death,} {\protect\JournalTitle{Proc. SPIE}} p. 790710 (2011).

\bibitem{Majno_Am_J_Pathol_1995}
G.~Majno and I.~Joris, \enquote{Apoptosis, oncosis, and necrosis. {An} overview
  of cell death.} {\protect\JournalTitle{The American Journal of Pathology}}
  \textbf{146}, 3--15 (1995).

\bibitem{Scholler2019OpEx}
J.~Scholler, \enquote{Motion artifact removal and signal enhancement to achieve
  in vivo dynamic full field oct,} {\protect\JournalTitle{Opt. Express}}
  \textbf{27}, 19562--19572 (2019).

\bibitem{Oldenburg_Optica2015}
A.~L. Oldenburg, X.~Yu, T.~Gilliss, O.~Alabi, R.~M. Taylor, and M.~A. Troester,
  \enquote{Inverse-power-law behavior of cellular motility reveals
  stromal\&\#x2013;epithelial cell interactions in {3D} co-culture by {OCT}
  fluctuation spectroscopy,} {\protect\JournalTitle{Optica}} \textbf{2},
  877--885 (2015). Publisher: Optical Society of America.

\bibitem{Leroux_BOE2016}
C.-E. Leroux, F.~Bertillot, O.~Thouvenin, and A.-C. Boccara,
  \enquote{Intracellular dynamics measurements with full field optical
  coherence tomography suggest hindering effect of actomyosin contractility on
  organelle transport,} {\protect\JournalTitle{Biomedical Optics Express}}
  \textbf{7}, 4501--4513 (2016). Publisher: Optical Society of America.

\end{thebibliography}

\clearpage
\section*{\Large Supplementary materials}

	Here we present the OCT intensity and cross sectional images of the spheroids pretested in the full-length manuscript.
	In addition, the measurement results of additional MCF-7 and HT-29 spheroids are presented and they are consistent with those presented in the full length manuscript.
	The tables elaborating the statistical significance analysis are also presented.

\section*{Supplementary Figures}
	\begin{figure}[hbpt!]
		\centering\includegraphics{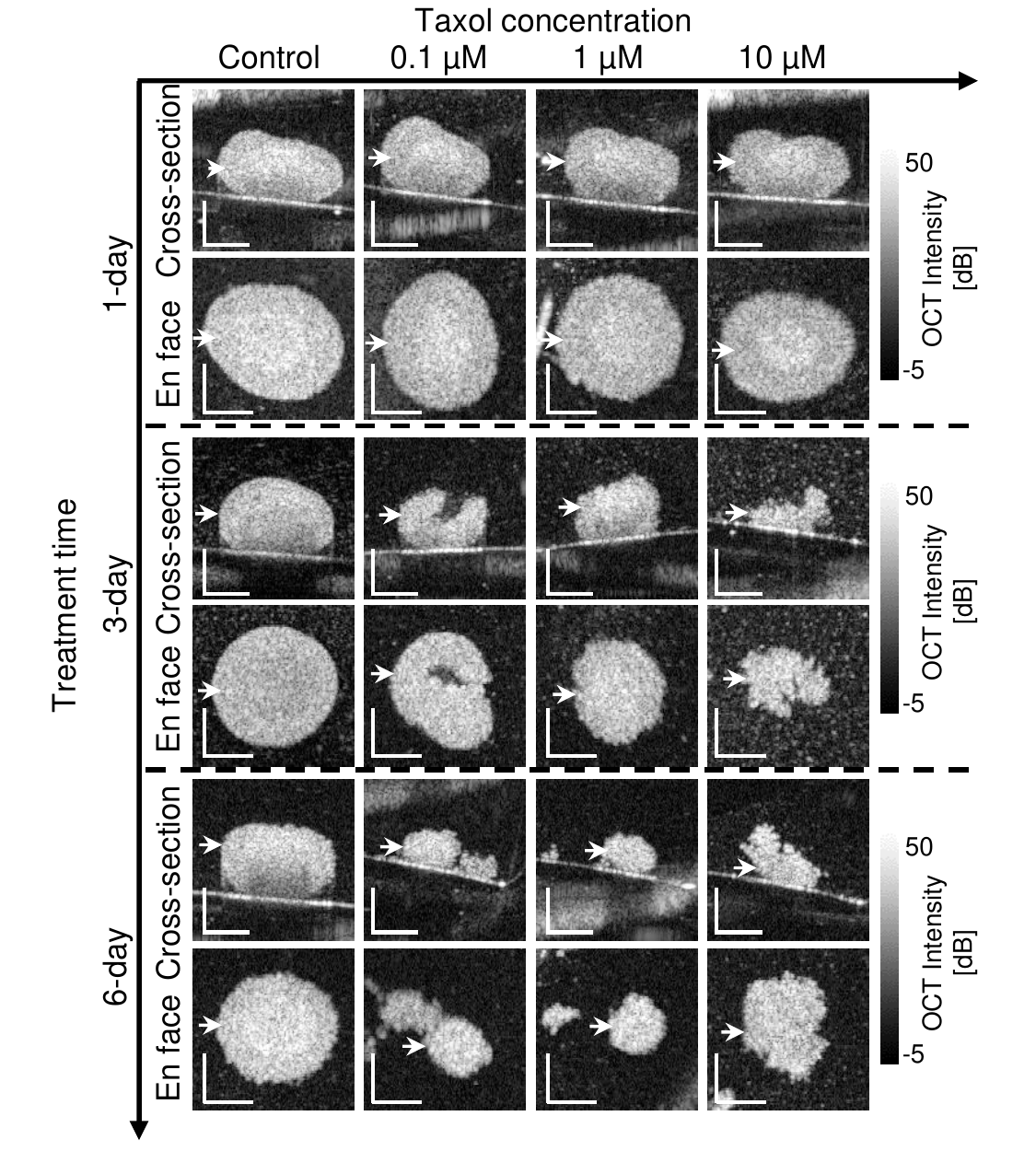} 
		\caption{%
			Cross-sectional and \enface OCT intensity images of the MCF-7 spheroid presented in Fig.\@ 3 in the full-length manuscript.
			The intensity images show the morphological  alteration of spheroid, while no tissue activity contrast can be observed.   
			Scale bars represent 200 \um.} 
		\label{fig:MCF7_Sphroid1-OCTInt}
	\end{figure}
	
	\clearpage
	
	\begin{figure}[hbpt!]
		\centering\includegraphics{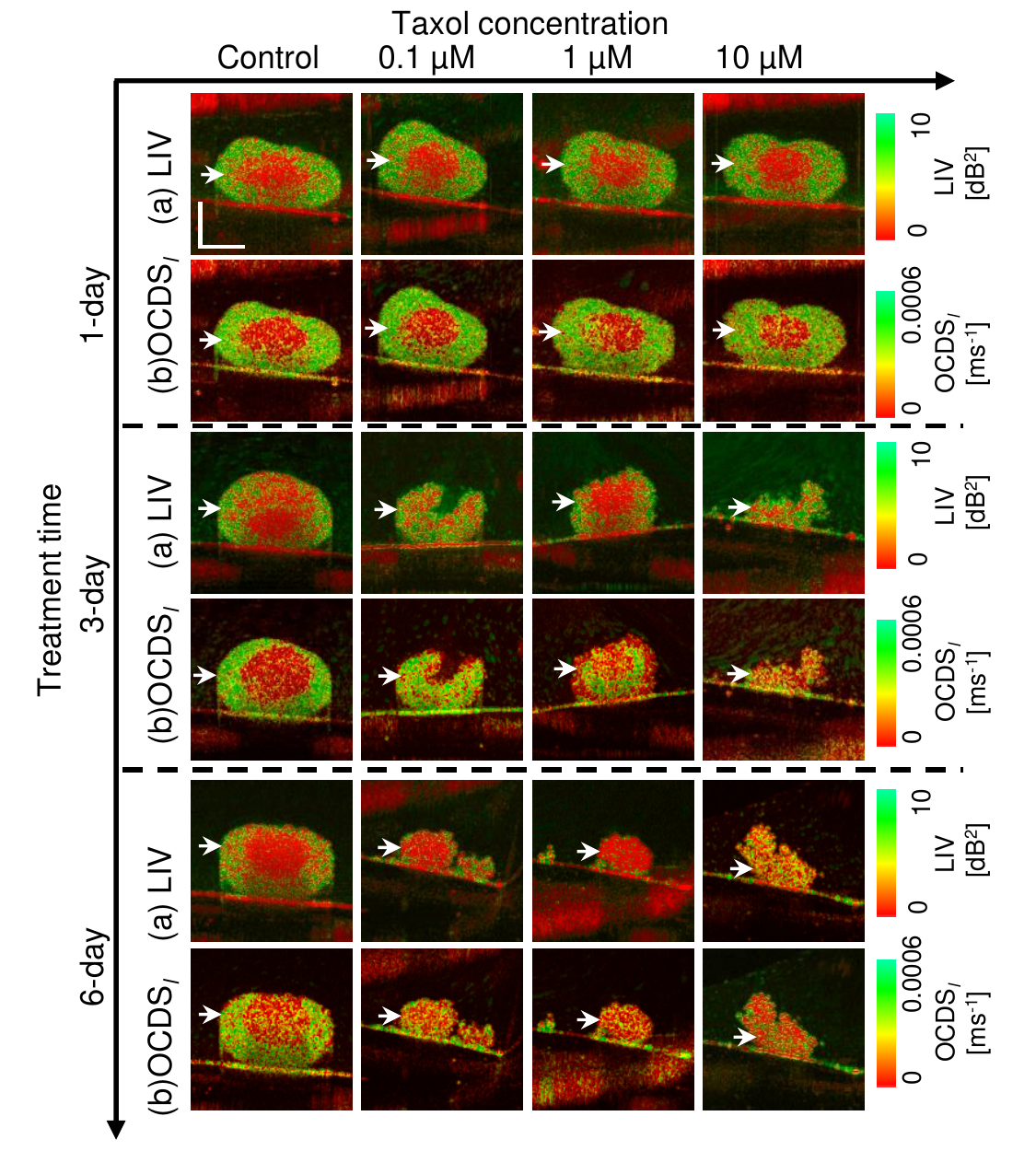} 
		\caption{Cross-sectional LIV and \OCDSl images of the MCF-7 spheroid treated with PTX.
			The images were extracted from the locations indicated by white arrow head on the \enface images in Fig.\@ 3 in the full length manuscript, and they show similar tendency to the \enface images.
			Scale bar represent 200 \um.}
		\label{fig:MCF7_Sphroid1-Cross-secImages}
	\end{figure}
	
	\clearpage

	\pagebreak
	\begin{figure}[hbpt!]
		\centering\includegraphics{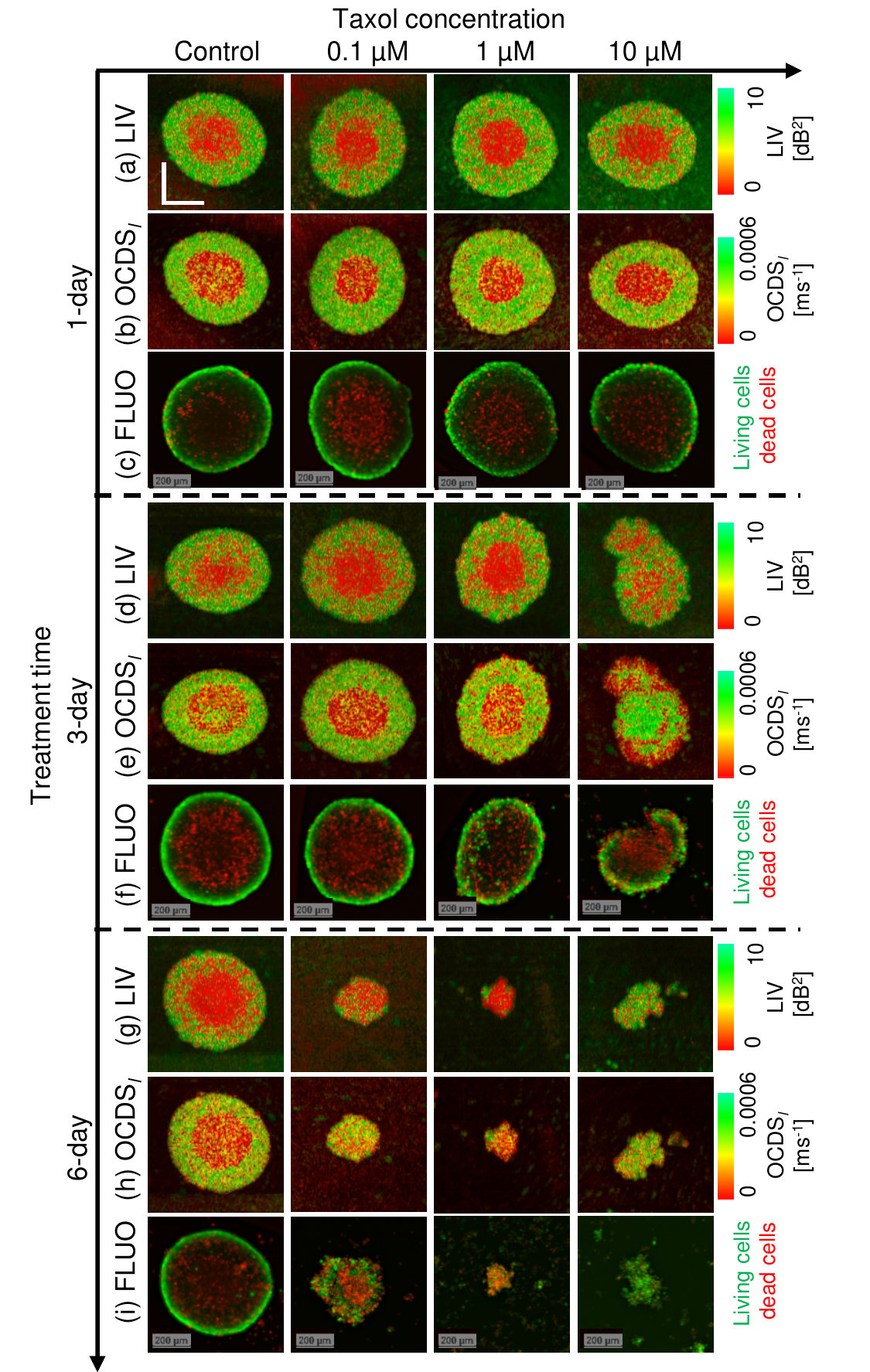} 
		\caption{\Enface LIV and \OCDSl images of additional MCF-7 spheroids treated with PTX.
			The measurement and treatment were performed with the same protocol to that of Fig.\@ 3 in the full-length manuscript.
			The images are presented in the same order as Fig.\@ 3 in the full-length manuscript. 
			Scale bar is applicable for all the LIV and \OCDSl images and it represents 200 \um.}
		\label{fig:MCF-7LIVSpheroid2}
	\end{figure}
	
	\pagebreak
	

	
	
	
	
	\clearpage
	
	
	\clearpage

	
	\begin{figure}[hbpt!]
		\centering\includegraphics{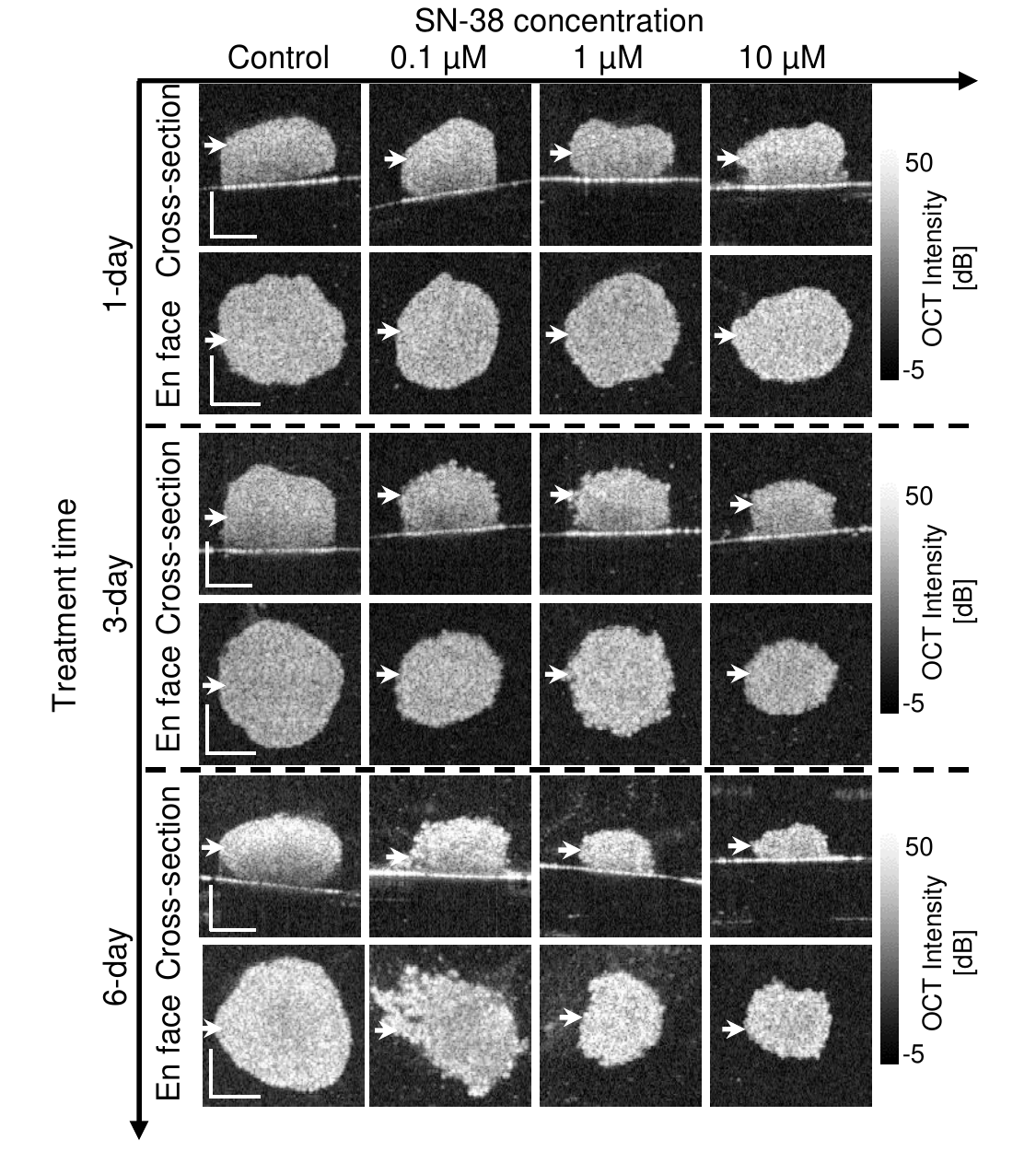} 
		\caption{Cross-sectional and \enface OCT intensity images of the HT-29 spheroids presented in Fig.\@ 6 in the full-length manuscript.
			The intensity images show the morphological  alteration of the HT-29 tumor spheroid.
			Scale bars represent 200 \um.}
		\label{fig:HT29_Sphroid1-OCTInt}
	\end{figure}
	
	\clearpage

	\begin{figure}[hbpt!]
		\centering\includegraphics{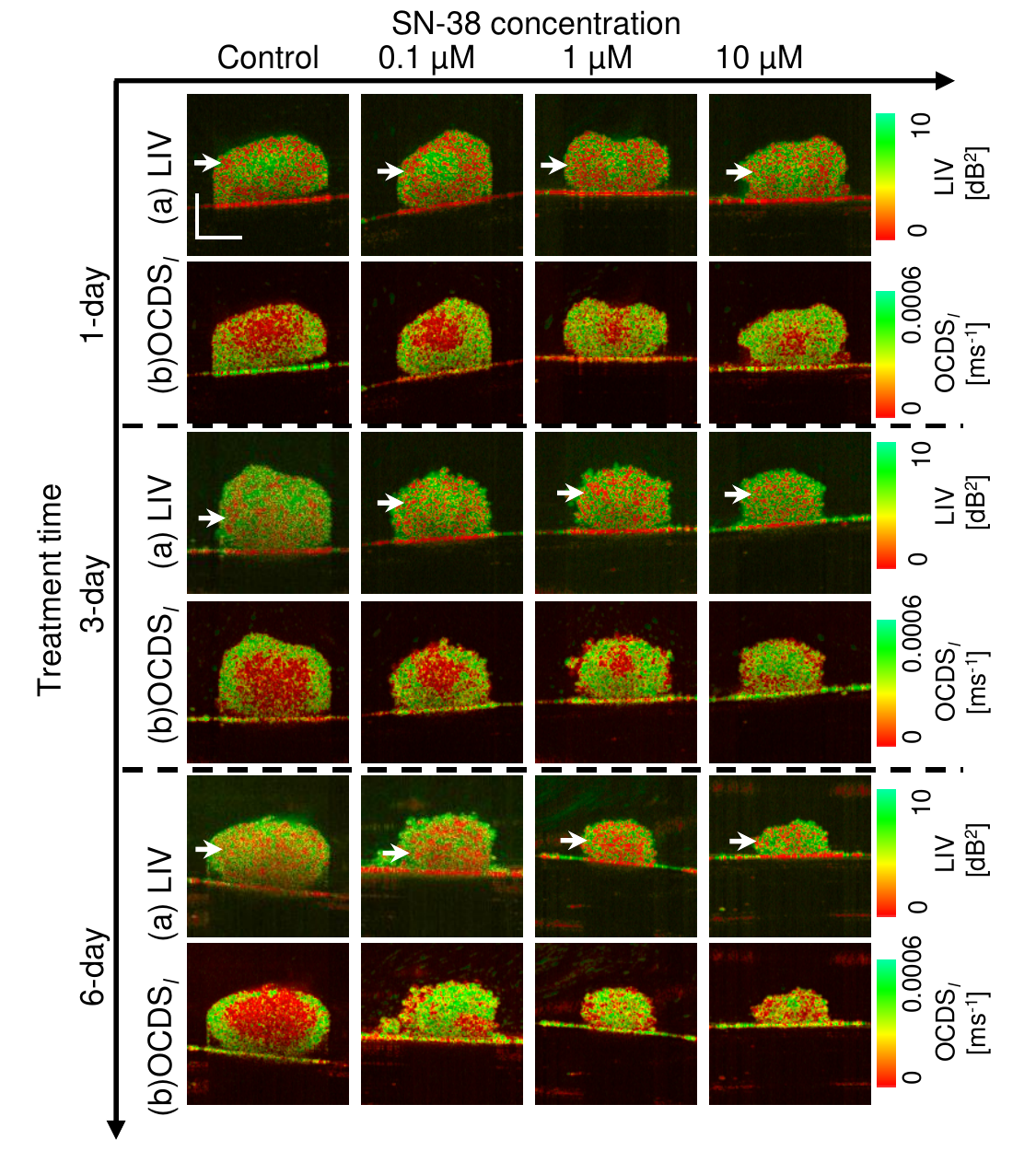} 
		\caption{The cross-sectional LIV and \OCDSl images  of HT-29 spheroids extracted from the locations indicated by white arrow heads in Fig.\@ 6 in the full length manuscript.
			The cross-section LIV and \OCDSl images show similar appearances to those of the \enface images pretested in Fig.\@ 6.
			Scale bar represent 200 \um.}
		\label{fig:HT29_Sphroid1-Cross-secImages}
	\end{figure}
	
	\clearpage
	
	\begin{figure}[hbpt!]
		\centering\includegraphics{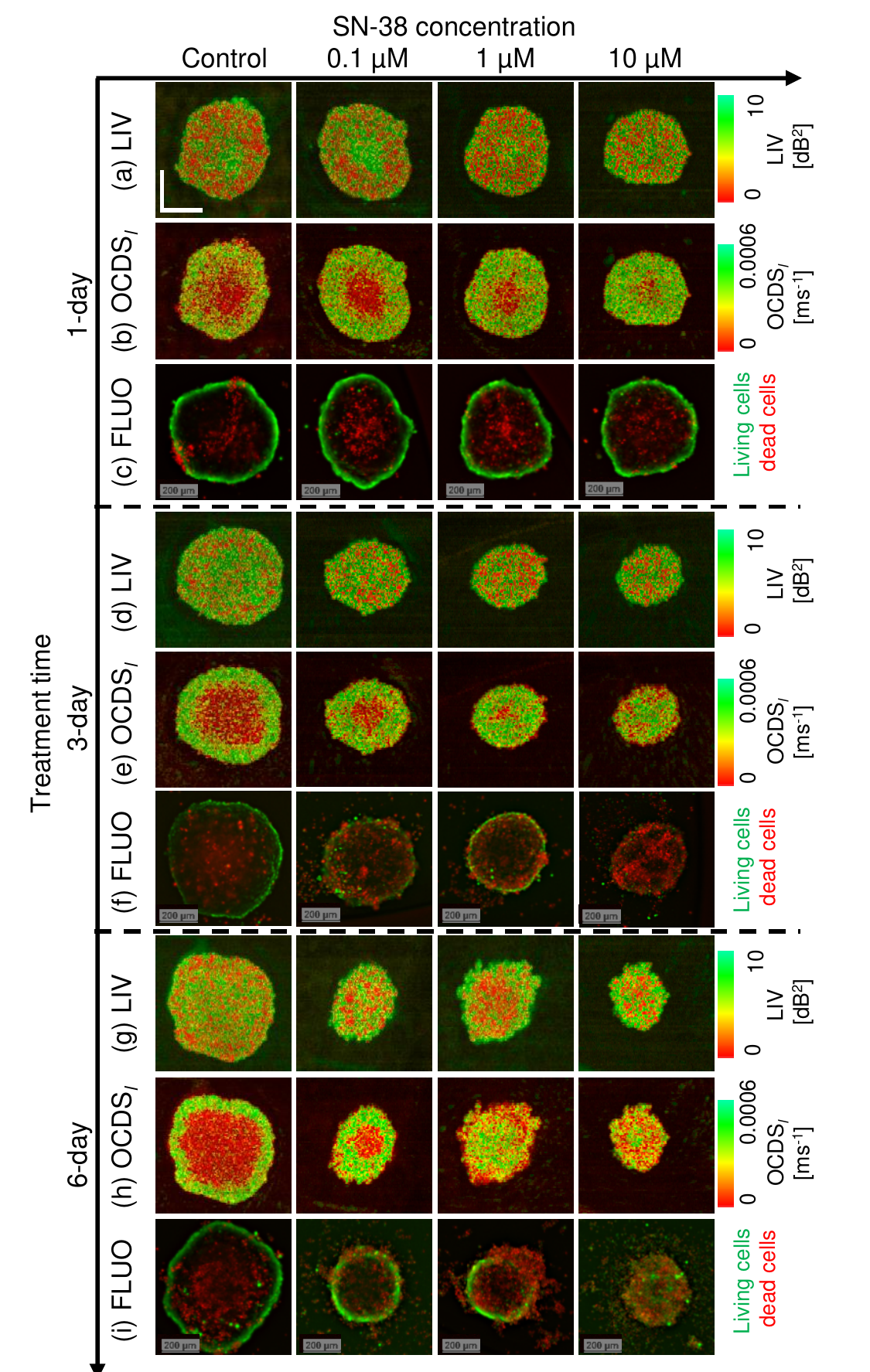} 
		\caption{The \enface LIV and \OCDSl images of additional HT-29 spheroids treated with SN-38.
			The images are presented in the same manner of Fig.\@ 6 in the full-length manuscript.
			The LIV and \OCDSl images show similar patterns to those presented in Fig.\@ 6 in the manuscript.
			Scale bar represents 200 \um.}
		\label{fig:HT-29LIVSpheroid5}
	\end{figure}
	
	\clearpage
	\section*{Supplementary Tables}
	\begin{table}[hbpt!]
		\centering
		\caption{Mann-Whitney significance test results of MCF-7 spheroid response to PTX based on the mean LIV, mean \OCDSl, LIV cut-off based necrotic cells ratio, and \OCDSl cut-off based necrotic cells ratio.
			The test is performed between each pair of treatment time points.} 
		\setlength{\tabcolsep}{3pt}
		\begin{tabular}{l|cccc} \hline \cellcolor{blue!25} {PTX concentration}
			& Control &  0.1-\uM & 1-\uM & 10-\uM \\ \hline 
			\cellcolor{green!25} \vtop{\hbox{\strut 1-day and 3-day}\hbox{\strut Mean LIV-based}\hbox{\strut [U-statistics, P-value]}}& [4.0,0.047]{*}&	[2.0,0.018]{*}&[0.0,0.006]{*}&[5.0,0.071] \\ \hline
			\cellcolor{green!25} \vtop{\hbox{\strut 3-day and 6-day}\hbox{\strut Mean LIV-based}\hbox{\strut [U-statistics, P-value]}}&[0.0,0.006]{*}&[0.0,0.006]{*}&[2.0,0.018]{*}&[5.0,0.071] \\ \hline
			\cellcolor{green!25} \vtop{\hbox{\strut 1-day and 6-day}\hbox{\strut Mean LIV-based}\hbox{\strut [U-statistics, P-value]}}&[0.0,0.006]{*}&[0.0,0.006]{*}&[0.0,0.006]{*}&[0.0,0.006]{*} \\ \hline	
			\cellcolor{yellow!25}\vtop{\hbox{\strut 1-day and 3-day}\hbox{\strut Mean \OCDSl-based}\hbox{\strut [U-statistics, P-value]}}& [3.0,0.030]{*}&[0.0,0.006]{*}&	[2.0,0.018]{*}&[0.0,0.006]{*} \\ \hline
			\cellcolor{yellow!25}\vtop{\hbox{\strut 3-day and 6-day}\hbox{\strut Mean \OCDSl-based}\hbox{\strut [U-statistics, P-value]}}&[0.0,0.006]{*}&[0.0,0.006]{*}&	[3.0,0.030]{*}&[7.0,0.148] \\\hline
			\cellcolor{yellow!25}\vtop{\hbox{\strut 1-day and 6-day}\hbox{\strut Mean \OCDSl-based}\hbox{\strut [U-statistics, P-value]}}&[0.0,0.006]{*}&[0.0,0.006]{*}&	[0.0,0.006]{*}&[0.0,0.006]{*} \\\hline
			
			\cellcolor{green!25} \vtop{\hbox{\strut 1-day and 3-day}\hbox{\strut For LIV-based necrotic cells ratio}\hbox{\strut [U-statistics, P-value]}}& [5.0,0.071]&	[3,0.030]{*}&	[0.0,0.006]{*}&	[4.0,0.047]{*} \\ \hline
			\cellcolor{green!25} \vtop{\hbox{\strut 3-day and 6-day}\hbox{\strut For LIV-based necrotic cells ratio}\hbox{\strut [U-statistics, P-value]}}&[0.0,0.006]{*}&	[0.0,0.006]{*}&	[4.0,0.047]{*}&	[3,0.030]{*} \\ \hline
			\cellcolor{green!25} \vtop{\hbox{\strut 1-day and 6-day}\hbox{\strut For LIV-based necrotic cells ratio}\hbox{\strut [U-statistics, P-value]}}&[0.0,0.006]{*}&	[0.0,0.006]{*}&	[0.0,0.006]{*}&	[0.0,0.006]{*}\\ \hline

			\cellcolor{yellow!25}\vtop{\hbox{\strut 1-day and 3-day}\hbox{\strut For \OCDSl-based necrotic cells ratio}\hbox{\strut [U-statistics, P-value]}}&[3.0,0.030]{*}&	[2.0,0.018]{*}&	[1.0, 0.010]{*}&	[0.0,0.006]{*} \\ \hline
			\cellcolor{yellow!25}\vtop{\hbox{\strut 3-day and 6-day}\hbox{\strut For \OCDSl-based necrotic cells ratio}\hbox{\strut [U-statistics, P-value]}}&[0.0,0.006]{*}&	[0.0,0.006]{*}&	[3.0,0.030]{*}&	[5.0,0.071]\\ \hline
			\cellcolor{yellow!25}\vtop{\hbox{\strut 1-day and 6-day}\hbox{\strut For \OCDSl-based necrotic cells ratio}\hbox{\strut [U-statistics, P-value]}}&[0.0,0.006]{*}&	[0.0,0.006]{*}&	[0.0,0.006]{*}&	[0.0,0.006]{*} \\ \hline
			
		\end{tabular}
		\label{tab:MCF7_TimepointsPairsStatistics}
	\end{table}
	
	\clearpage
	

	
	
	
	
	\begin{table}[hbpt!]
		\centering
		\caption{%
			Mann-Whitney significance test results of HT-29 spheroid response to SN-38 based on the mean LIV, mean \OCDSl, LIV cut-off based necrotic cells ratio, and \OCDSl cut-off based necrotic cells ratio.
			The test was done between each pair of treatment time points.} 
		\setlength{\tabcolsep}{3pt}
		\begin{tabular}{l|cccc} \hline \cellcolor{blue!25} {SN-38 concentration}
			& Control &  0.1-\uM & 1-\uM & 10-\uM \\ \hline 
			\cellcolor{green!25} \vtop{\hbox{\strut 1-day and 3-day}\hbox{\strut Mean LIV-based}\hbox{\strut [U-statistics, P-value]}}& [5.0,0.071]&	[7,0.148]&[5.0, 0.071]&[1.0, 0.010]{*} \\ \hline
			\cellcolor{green!25} \vtop{\hbox{\strut 3-day and 6-day}\hbox{\strut Mean LIV-based}\hbox{\strut [U-statistics, P-value]}}&[0.0, 0.006]{*}&[9,0.265]&	[8,0.201]&[5.0, 0.135] \\ \hline
			\cellcolor{green!25} \vtop{\hbox{\strut 1-day and 6-day}\hbox{\strut Mean LIV-based}\hbox{\strut [U-statistics, P-value]}}&[11,0.417]&[11, 0.417]&[6.0,0.105]&	[3,0.055] \\ \hline	
			\cellcolor{yellow!25}\vtop{\hbox{\strut 1-day and 3-day}\hbox{\strut Mean \OCDSl-based}\hbox{\strut [U-statistics, P-value]}}& [9,0.265]&[6,0.105]&	[8,0.201]&[2,0.018]{*} \\ \hline
			\cellcolor{yellow!25}\vtop{\hbox{\strut 3-day and 6-day}\hbox{\strut Mean \OCDSl-based}\hbox{\strut [U-statistics, P-value]}}&[6,0.105]&[7,0.148]&	[0.0,0.006]{*}&[3,0.055] \\ \hline
			\cellcolor{yellow!25}\vtop{\hbox{\strut 1-day and 6-day}\hbox{\strut Mean \OCDSl-based}\hbox{\strut [U-statistics, P-value]}}&[10,0.338]&[10, 0.338]&	[0.0,0.006]{*}&[0.0,0.009]{*} \\ \hline
			\cellcolor{green!25} \vtop{\hbox{\strut 1-day and 3-day}\hbox{\strut For LIV-based necrotic cells ratio}\hbox{\strut [U-statistics, P-value]}}& [5.0,0.071]&	[4.0,0.047]{*}&	[5.0, 0.071]&[1.0,0.010]{*} \\ \hline
			\cellcolor{green!25} \vtop{\hbox{\strut 3-day and 6-day}\hbox{\strut For LIV-based necrotic cells ratio}\hbox{\strut [U-statistics, P-value]}}&[7.0, 0.148]&[10.0,0.338]&[5.0,0.0718]&[8.0,0.356] \\ \hline
			\cellcolor{green!25} \vtop{\hbox{\strut 1-day and 6-day}\hbox{\strut For LIV-based necrotic cells ratio}\hbox{\strut [U-statistics, P-value]}}&[5.0, 0.0718]&[6.0,0.105]&[4.0,0.047]{*}&[1.0,0.018]{*}\\ \hline
			\cellcolor{yellow!25}\vtop{\hbox{\strut 1-day and 3-day}\hbox{\strut For \OCDSl-based necrotic cells ratio}\hbox{\strut [U-statistics, P-value]}}& [9.0,0.265]&[10,0.338]&	[9,0.265]&[0.0,0.006]{*}\\ \hline
			\cellcolor{yellow!25}\vtop{\hbox{\strut 3-day and 6-day}\hbox{\strut For \OCDSl-based necrotic cells ratio}\hbox{\strut [U-statistics, P-value]}}&[6.0,0.105]&[7.0,0.148]&	[0.0,0.006]{*}&[4.0,0.088]\\ \hline
			\cellcolor{yellow!25}\vtop{\hbox{\strut 1-day and 6-day}\hbox{\strut For \OCDSl-based necrotic cells ratio}\hbox{\strut [U-statistics, P-value]}}&[8.0,0.201]&[10.0,0.338]&	[0.0,0.006]{*}&[0.0,0.009]{*} \\ \hline
		\end{tabular}
		\label{tab:HT29_TimepointsPairsStatistics}
	\end{table}
	
		

\end{document}